\documentclass[journal=jacsat,manuscript=article]{achemso}

\usepackage{setspace}
\usepackage{xcolor}
\usepackage{graphicx}
\usepackage{academicons}
\usepackage{orcidlink}
\usepackage{multirow}
\usepackage{chemformula}
\usepackage{makecell}
\usepackage{colortbl}
\usepackage{enumitem}
\usepackage{placeins}
\usepackage{pifont}

\usepackage{xr-hyper}
\usepackage{hyperref}
 
\title[An \textsf{achemso} demo]
  {Redox-Active Halide Materials for Cathode Applications \\ \vspace{5mm} \small \today}

\author{Zhuohan Li}
\affiliation{Materials Sciences Division, Lawrence Berkeley National Laboratory, Berkeley, California, United States}
\author{Gerbrand Ceder}
\affiliation{Materials Sciences Division, Lawrence Berkeley National Laboratory, Berkeley, California, United States}
\alsoaffiliation{Department of Materials Science and Engineering, University of California, Berkeley, California, United States}
\email{gceder@berkeley.edu}

\abbreviations{IR,NMR,UV}

\date{\today}

\begin{document}







\begin{abstract}

Electrochemically redox-active halide (eREAL) materials are an emerging class of materials that combine high Li-ion conductivity with transition-metal redox activity, making them promising candidates for cathode or catholyte applications. As a redox-active catholyte, they could significantly increase the energy density of solid-state batteries. In this work, we perform first-principles calculations on Li–M–Cl (M = 3\textit{d} transition metals) ternaries to establish such a theoretical foundation for their stability and electrochemical activity. We map the phase stability of eREAL structures with varying metal-to-Cl ratio, transition-metal species, oxidation states, and anion frameworks, and compute cation and anion redox potentials. We find that the high ionicity of metal--Cl bonds elevates cation redox potentials above those of conventional oxide cathodes, but also will promote Cl oxidation and Cl–Cl dimerization at high voltages, which may limit the stability of these materials. Anion substitution effectively tunes both cation and anion redox potentials, with F substitution standing out as a viable route to extend the reversible voltage window. Beyond the anion redox issue, eREAL compounds generally exhibit flat voltage profiles, which potentially poses an electrochemical compatibility challenge when paired with active materials that operate at different voltage values or over wider voltage ranges. Collectively, our study provides a comprehensive analysis for redox behavior of eREAL materials, paving the way for their rational design and optimization in next-generation battery applications.

\end{abstract}

\section{Introduction}

The realization of practical solid-state batteries is constrained by the lack of a single type of solid electrolyte that meets all key battery performance requirements \cite{Kovalenko2025_catholyte, Ko2026_review_practical_SSE}. Oxide-based electrolytes are generally too stiff and brittle to maintain effective interfacial contact, which is essential for low interfacial resistance and long-term cycling stability, with commercially available cathode materials such as \ch{LiNi_xMn_yCo_{1-x-y}O2} and \ch{LiFePO4}. Although sulfide electrolytes offer superior ionic conductivity and mechanical plasticity, they suffer from poor electrochemical and moisture stability. Halide systems are increasingly attracting attention for their balanced properties, combining high ionic conductivity and mechanical deformability of sulfides with high oxidation stability of oxides. In particular, the relatively high oxidative stability of halides makes them promising for integration within a composite cathode as catholytes.

Electronic conductivity is usually suppressed in solid-state electrolytes so that they can function as a separator between cathode and anode. However, when used as a catholyte in a composite cathode, the low electronic conductivity of the solid electrolyte (and sometimes active material) necessitates the incorporation of additional electron-conducting materials such as carbon additives. A typical composite cathode requires over $\sim$40 vol$\%$ to be filled with these redox-inactive components (liquid or solid electrolytes and carbon additives), which reduces the volumetric energy density \cite{Shearing2021_porousity, Shi2020_SSB_porousity,Gao2022_solid_state_cathode}. In solid-state batteries,, a high electrolyte-to-active material ratio is often used to improve rate capability and cycle life, as it can compensate for the loss of contact between the cathode and solid electrolyte arising from the repeated expansion and contraction of the cathode upon cycling. \cite{Ko2026_review_practical_SSE}.

To address these limitations, the use of electrochemically redox-active halide (eREAL) materials as catholytes has been proposed as a novel concept \cite{Song2024_LVC,Zhang2024_LFZC}. The eREAL materials incorporate redox-active transition metals into halide-based solid electrolytes to create mixed ionic-electronic conductivity and to add extra Li capacity. Beyond catholyte applications, eREAL materials have also been evaluated as stand-alone cathodes \cite{Wang2023_LTC,Liu2024_FeCl3,Fu2025_all_in_one}.

Given the promise of this class of materials, we map out in this paper the basic stability and electrochemical behavior of these materials. In this work, we perform first-principles calculations to evaluate chloride-based eREAL materials for cathode applications. Using first-principles calculations, we find that for most Li--M--Cl ratios we investigated, a ccp Cl anion framework is preferred over an hcp framework. The cation and anion redox activity is probed by topotactically delithiating these ccp structures and compared to conventional transition-metal oxide cathodes. Our results indicate that the high ionicity of metal--Cl bonds systematically raises the cation redox potentials in chlorides relative to oxides, but also promotes Cl oxidation and Cl--Cl dimerization at highly charged states, which may become a limiting factor for their application. The low metal-to-anion ratio in chlorides further limits the hybridization that protects \ch{Cl-} orbitals from oxidation. We also investigate the effect of anion substitution on mitigating the anion oxidation issue. Electrochemical and mechanical compatibilities of eREAL materials in a composite cathode are also discussed. Our study aims to understand the strength and limitation of eREAL materials, offering a blueprint for further engineering them for high-performance cathode applications.

\section{Result}
\subsection{Phase stability in Li--M--Cl ternary chloride systems}

\begin{figure*}[h]
\centering
\includegraphics[width=\linewidth]{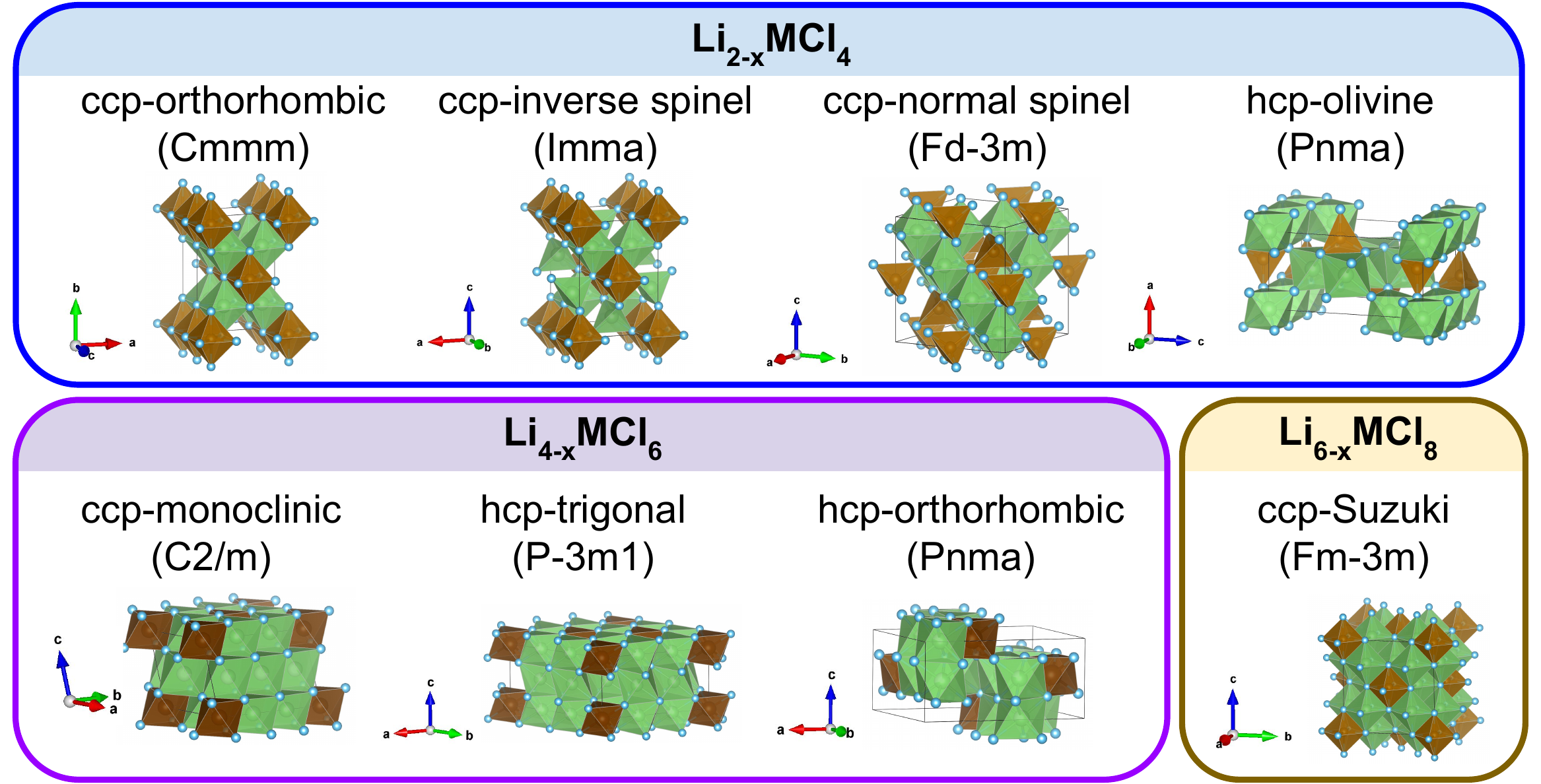}
\caption{Schematics of experimentally known close-packed ternary chloride crystal structures with varying metal-to-Cl ratios, i.e., \ch{Li_{2-x}MCl4}, \ch{Li_{4-x}MCl4}, and \ch{Li_{6-x}MCl4}, where M stands for non-Li metal cations. Li, Cl, and metal cations are colored with green, light blue, and brown, respectively.}
\label{fig:schematic_crystals} 
\end{figure*}

To broadly evaluate the potential of chlorides as redox-active conductors, we consider three representative chloride compositions: \ch{Li_{2-x}MCl4}, \ch{Li_{4-x}MCl6}, and \ch{Li_{6-x}MCl8}, where M stands for 3\textit{d} transition metals. These form as a set of compounds with decreasing metal-to-Cl ratio. Most experimentally reported crystalline halide-type ionic conductors adopt either cubic close-packed (ccp) or hexagonal close-packed (hcp) anion frameworks \cite{Li2020_halide_review, He2023_halide_review}. For example, chlorides with divalent metal cations typically crystallize in ccp-derived structures, such as the orthorhombic (Cmmm) and inverse spinel (Imma) \ch{Li2MCl4} phases, and the Suzuki-type \ch{Li6MCl8} phase (Fm$\overline{3}$m). An exception is \ch{Li2ZnCl4}, which adopts the olivine structure with an hcp Cl anion framework (Pnma) at elevated temperatures ($>500$ K) but crystallizes in a ccp normal spinel structure (Fd$\overline{3}$m) at low temperatures. For \ch{Li3MCl6} compositions, high Li-ion conductivity has been reported for both ccp (monoclinic, C2/m) and hcp-based phases (trigonal, P$\overline{3}$m1, or orthorhombic, Pnma). These close-packed chloride structures are illustrated in Figure \ref{fig:schematic_crystals}.  

\FloatBarrier

We identify the lowest-energy structure for a range of compositions spanning varying metal-to-Cl ratios, 3\textit{d} transition-metal species, oxidation states, and anion frameworks. The phase stability of different structures are evaluated by calculating the energy above the convex hull ($E_{\mathrm{hull}}$) with respect to all r$^2$SCAN-calculated structures that are present in the same chemical space in the Materials Project \cite{Jain2013_MP, Horton2025_MP}, as well as other chemically relevant structures calculated in this study. For those structures where multiple Li/vacancy arrangements are possible, we enumerate different Li/vacancy arrangements using the method described in Section SX in the Supporting Information. Unless noted otherwise, the r$^2$SCAN functional is used for all DFT calculations, as it better reproduces experimental voltage values than the Perdew--Burke--Ernzerhof (PBE) functional \cite{Isaacs2020_SCAN_benchmark}, and is appropriate here due to the short-range Van der Waals interaction between Cl anions \cite{Yang2019_SCAN_benchmark}. 

\begin{figure*}[t]
\centering
\includegraphics[width=0.7\linewidth]{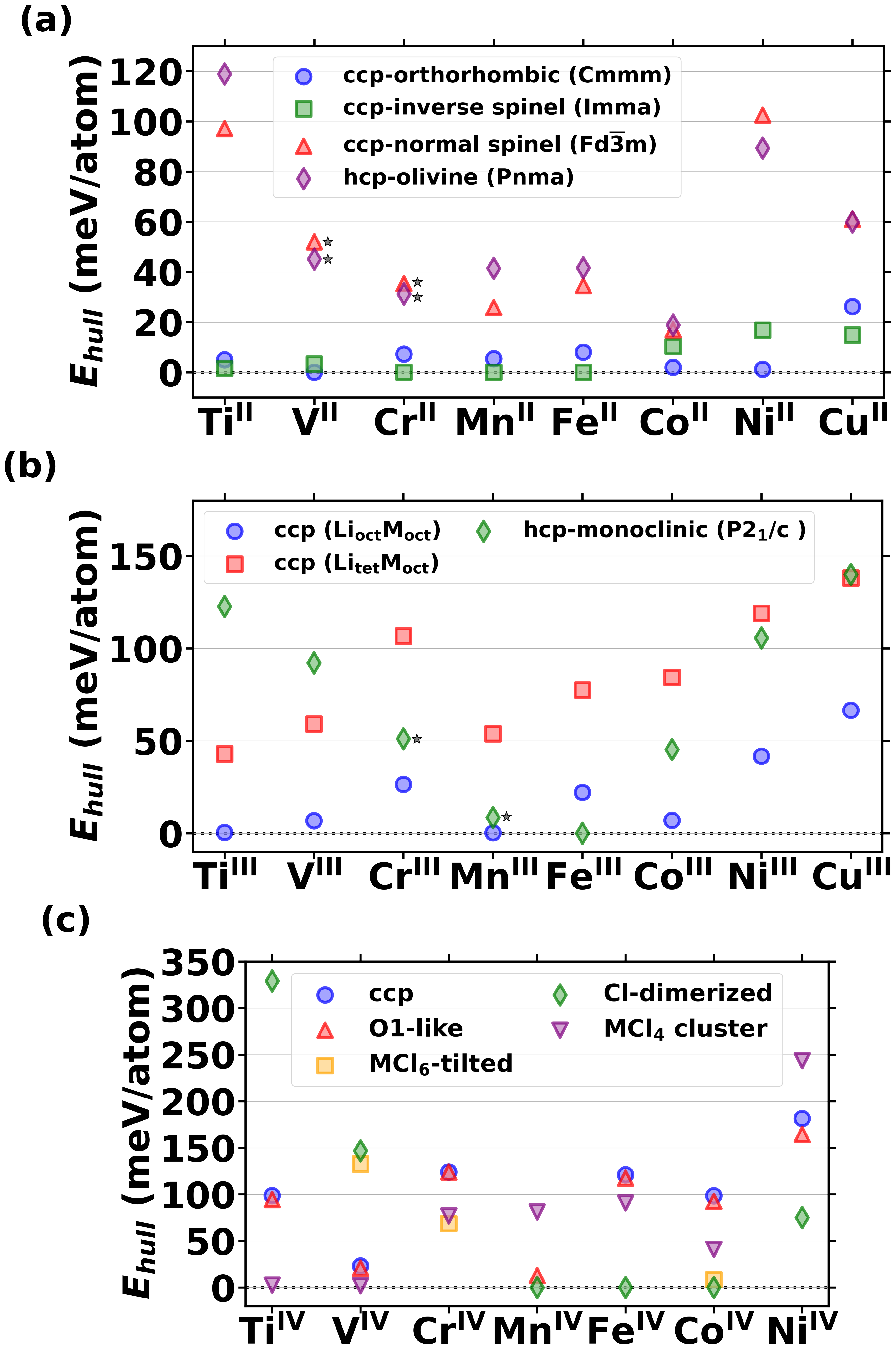}
\caption{$E_{\mathrm{hull}}$ of (a) \ch{Li2M^{II}Cl4}, (b) \ch{LiM^{III}Cl4}, and (c) \ch{M^{IV}Cl4} across different polymorphs and 3\textit{d} transition-metal cations. The black starts in (a) and (b) indicate that structures that undergo large structural distortions with transition-metal displacements during relaxations, as described in detail in the main text. The O1-like structure and the other three types of structural distortions (i.e., \ch{MCl6}-tilted, Cl-dimerized, and \ch{MCl4} cluster) at IV formal oxidation state are explicitly labeled in (c). }
\label{fig:LixMCl4_Ehull} 
\end{figure*}
\FloatBarrier

Figure \ref{fig:LixMCl4_Ehull} (a) shows the calculated $E_{\mathrm{hull}}$ values of \ch{Li2M^{II}Cl4} for four polymorphs, i.e., orthorhombic, inverse spinel, normal spinel, and olivine. The orthorhombic and inverse spinel phases share the same transition-metal cation arrangement within the ccp anion lattice but differ primarily in Li site occupancy. In the orthorhombic phase, all Li ions occupy octahedral sites, making this essentially a cation-deficient ordered rocksalt, whereas in the inverse spinel, half of the Li ions occupy tetrahedral sites. The transition-metal cations reside in octahedral sites in these two polymorphs, but are tetrahedrally coordinated in the normal spinel and olivine structures. Some structures distort and lower the symmetry upon relaxation (see Figures SX in the Supporting Information for the distorted crystal structures). For example, both \ch{Li2CrCl4} and \ch{Li2VCl4} exhibit large distortions in their normal spinel and olivine structures by displacing the transition-metal cations from the tetrahedral centers so that \ch{Cr^{II}} and \ch{V^{II}} cations are coordinated by 5 and 6 Cl anions, respectively, instead of 4. In the distorted normal spinel and distorted olivine structures of \ch{Li2VCl4}, the displacement of \ch{V^{II}} is accompanied with the migration of some Li ions from the octahedral to tetrahedral sites. As can be seen from Figure \ref{fig:LixMCl4_Ehull} (a), for all transition-metal cations considered here, the ccp structures with octahedrally coordinated transition metals (the orthorhombic and inverse spinel phases) have lower energies than those with tetrahedrally coordinated transition metals (the normal spinel and olivine phases) or their distorted derivatives. The selection between the orthorhombic and inverse spinel phases depends on the transition-metal cations, although their energies differ by less than 20 meV/atom, suggesting that Li site energy difference between the octahedral and tetrahedral sites is less than 0.14 eV per Li.

Figure \ref{fig:LixMCl4_Ehull} (b) shows the $E_{\mathrm{hull}}$ of \ch{LiM^{III}Cl4} structures, which are generated by topotactically extracting Li from the parent \ch{Li2M^{II}Cl4} structures. Because the orthorhombic and inverse spinel phases consistently have the lowest energies for \ch{Li2M^{II}Cl4}, we perform Li extraction only from these two polymorphs. For the inverse spinel phase, we consider two delithiated configurations: one where all the tetrahedral Li are extracted (leaving only octahedral Li in the structrue), and another where all the octahedral Li are extracted (leaving only tetrahedral Li in the structrue). An octahedral-Li-only configuration can also be obtained by delithiating the orthorhombic phase, where we remove half of the octahedral Li from the parent orthorhombic \ch{Li2M^{II}Cl4} structure and enumerate different Li/vacancy arrangements. Our result in Figure \ref{fig:LixMCl4_Ehull} (b) shows that structures with all Li in octahedral sites (denoted "\ch{Li_{$\mathrm{oct}$}M_{$\mathrm{oct}$}}" in the figure) have lower energy than those with all Li in tetrahedral sites (denoted "\ch{Li_{$\mathrm{tet}$}M_{$\mathrm{oct}$}}" in the figure). This suggests that, when tetrahedral Li is present in \ch{Li2M^{II}Cl4}, it is preferentially extracted during delithiation, which is in contrast to Li stability in oxide spinels. Consequently, both Imma and Cmmm structures will be delithiated to the same \ch{Li_{$\mathrm{oct}$}M_{$\mathrm{oct}$}} configuration at the III formal oxidation state (\ch{LiM^{III}Cl4}). Figure \ref{fig:LixMCl4_Ehull} (b) shows that the energy difference between the \ch{Li_{$\mathrm{tet}$}M_{$\mathrm{oct}$}} and \ch{Li_{$\mathrm{oct}$}M_{$\mathrm{oct}$}} configurations ranges from 40 to 80 meV/atom,  i.e., 0.24–0.48 eV per Li, exceeding the Li site energy difference at the II formal oxidation state ($<0.14$ eV per Li, see Figure \ref{fig:LixMCl4_Ehull} (a)). This indicates that tetrahedral Li is destabilized in structures with a high concentration of Li vacancies. In addition to the topotactically delithiated structures, we also consider a monoclinic (P$2_1$/c) phase which is experimentally observed for \ch{LiFeCl4} \cite{Zhang2024_LFZC} (schematically shown in Figure SX in the Supporting Information). In this phase, Li and Fe occupy octahedral and tetrahedral sites, respectively, within the hcp Cl anion lattice. Upon relaxation from this monoclinic starting structure, \ch{LiMn^{III}Cl4} and \ch{LiCr^{III}Cl4} converge to distorted structures with the transition metals displaced from the tetrahedral to octahedral sites. In addition, Li ions migrate from the octahedral to tetrahedral sites in the distorted monoclinic \ch{LiCr^{III}Cl4} structure. All other monoclinic \ch{LiMCl4} structures with transition metals other than Mn and Cr retain the undistorted hcp-monoclinic symmetry upon relaxation. Figure \ref{fig:LixMCl4_Ehull}(b) shows that the ccp structure with both Li ions and transition-metal cations in octahedral sites (\ch{Li_{$\mathrm{oct}$}M_{$\mathrm{oct}$}}) has the lowest energy for all transition metals except Fe, for which the hcp-monoclinic phase is most stable, in agreement with the experiment. The ccp and hcp-monoclinic \ch{LiFeCl4} polymorphs differ in energy by 22 meV/atom. This energy difference indicates that \ch{Li2FeCl4} may be susceptible to a non-topotactic phase transformation upon delithiation (i.e., a conversion-type behavior), which would reduce the average voltage of the II/III redox couple by 0.13 V relative to the purely topotactic delithiation pathway. This conversion is not certain considering that almost all oxide cathodes have a driving force for non-topotactic conversion in their delithiation state, but are nonetheless metastable so that they can be cycled well.

\begin{figure*}[h]
\centering
\includegraphics[width=0.8\linewidth]{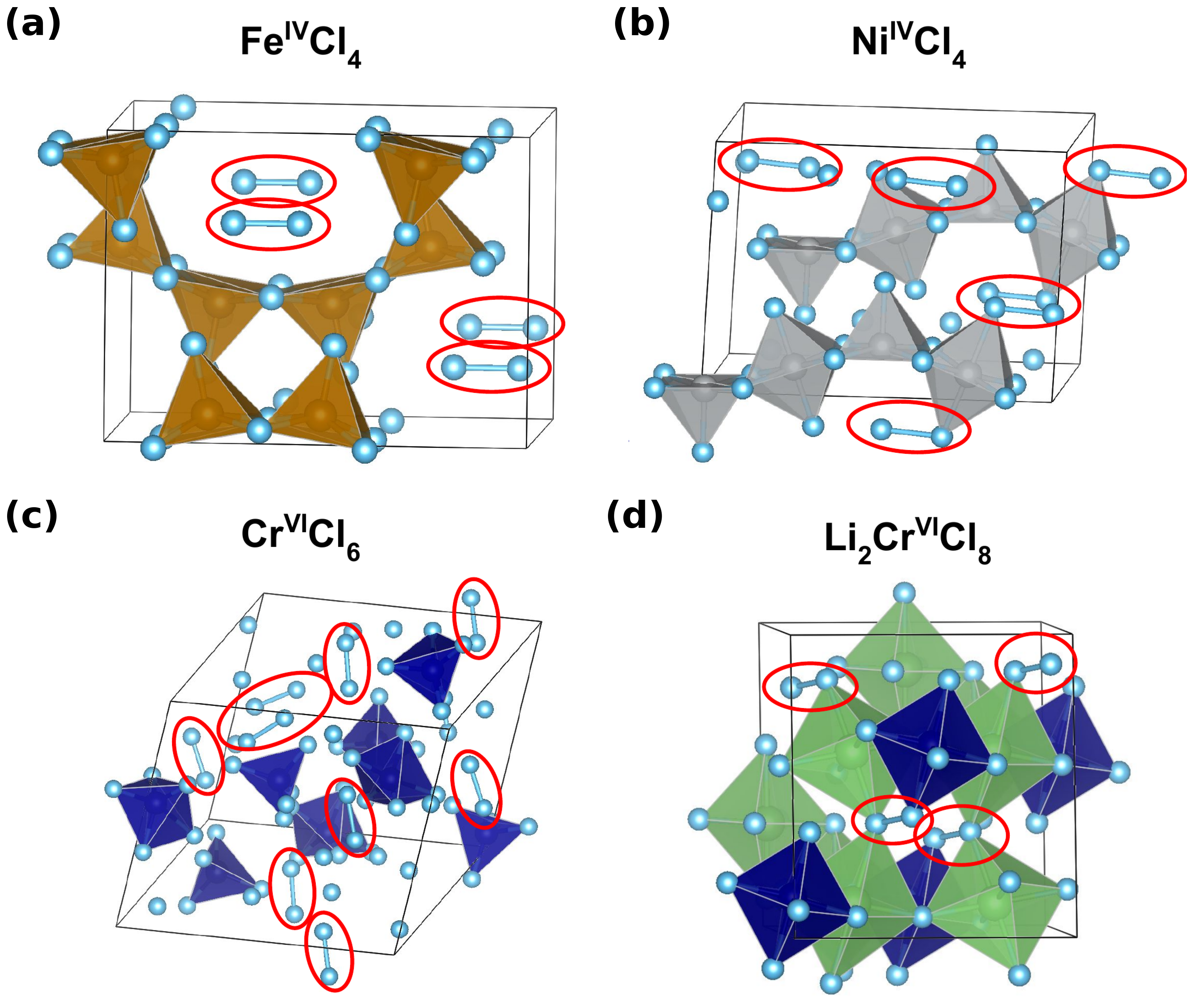}
\caption{Relaxed structures of (a) \ch{Cr^{VI}Cl6} and (b)  \ch{LiCr^{VI}Cl8}. In \ch{Cr^{VI}Cl6}, the bulk structure decomposes into \ch{Cl2} molecules and [\ch{CrCl5}] or [\ch{CrCl4}] clusters. In \ch{Li2Cr^{VI}Cl8}, Cl--Cl dimerization is observed with large structural distortions away from the original ccp anion lattice. Li, Cr, and Cl are colored by green, dark blue, and light blue, respectively. The dimerized Cl anions are highlighted with red circles.}
\label{fig:Cl_dimerization} 
\end{figure*}

In Figure \ref{fig:LixMCl4_Ehull}(c), we show $E_{\mathrm{hull}}$ at the IV formal oxidation state, i.e., \ch{M^{IV}Cl4}. All Li ions have been extracted in this composition. Our results show that for these topotactically delithiated ccp structures, $E_{\mathrm{hull}}$ values are higher than 100 meV/atom for most transition metals, indicating that topotactic delithiation is in general unstable at this high oxidation state. Notably, the \ch{Mn^{IV}Cl4} showed a O3-to-O1-like transition through spontaneous in-plane gliding during structural relaxation (schematically shown in Figure SX in the Supporting Information). The energy of O1-like structures substituted with other transition metals are also calculated. Our results indicate that the O1-like structures are generally lower in energy than O3-like structures, but the energy difference is generally less than 6 meV/atom (0.03 eV/f.u.) for \ch{Ti^{IV}}, \ch{V^{IV}}, \ch{Cr^{IV}}, \ch{Fe^{IV}}, and \ch{Co^{IV}}, and is at most 17 meV/atom (0.085 eV/f.u.) for \ch{Ni^{IV}}. In oxides, such O3-O1 transitions are occasionally observed and appear fully reversible \cite{Aydinol1998_LCO,Ruff2023_O3_O1_NMC}. 

We observe several types of large structural distortions are possible when the close-packed symmetry is allowed to break at this high oxidation state. For example, the starting ccp \ch{TiCl4} structure relaxes into isolated [\ch{TiCl4}] tetrahedral clusters (denoted as "\ch{MCl4} cluster" in Figure \ref{fig:LixMCl4_Ehull}(c)), while the structures with other transition metals undergo distortions that lead to either Cl--Cl dimerization (denoted as "Cl-dimerized" in Figure \ref{fig:LixMCl4_Ehull}(c)) or tilting of [\ch{MCl6}] octahedra with respect to edge-sharing octahedra (denoted as "\ch{MCl6}-tilted" in Figure \ref{fig:LixMCl4_Ehull}(c)). The Cl-dimerized \ch{FeCl4} and \ch{NiCl4} structures are schematically shown in Figure \ref{fig:Cl_dimerization} (a) and (b), respectively, while the other two types of highly distorted structures (i.e., the "\ch{MCl4} cluster" and "\ch{MCl6}-tilted" structures) are shown in Figure SX in the Supporting Information. To determine the lowest-energy structure for each transition metal, we further estimate the energies of these distorted structures for all the 3\textit{d} transition metals. We obtain Cl-dimerized structures after relaxation for all substituted metals except for \ch{Cr^{IV}}, for which the structure instead spontaneously relaxes into a \ch{CrCl6}-tilted structure. Among the substituted \ch{MCl6}-tilted structures, only those substituted with \ch{V^{IV}}, \ch{Cr^{IV}}, and \ch{Co^{IV}} retain the [\ch{MCl6}]-tilted motif after relaxation, whereas those for the other transition metals relax into either \ch{MCl4} cluster or Cl-dimerized structures. Figure \ref{fig:LixMCl4_Ehull}(c) shows that these highly distorted structures have a lower energy than the topotactically delithiated ccp or O1-like structures for all transition-metal species, while the lowest-energy distortion depends on the transition metal. The earlier transition metals, i.e., Ti and V, favor the formation of \ch{MCl4} clusters, whereas Cl--Cl dimerization lowers the energy more for transition metals later than Cr. The lowest-energy \ch{CrCl4} structure exhibit tilting of [\ch{CrCl6}] octahedra without Cl--Cl dimerization. These structural distortions lower the energy by 55–121 meV/atom (0.275–0.605 eV/f.u.) relative to the topotactically delithiated ccp structure for \ch{Ti^{IV}}, \ch{Cr^{IV}}, \ch{Fe^{IV}}, \ch{Co^{IV}}, and \ch{Ni^{IV}}. In contrast, the energy reduction from distortion is much smaller for \ch{V^{IV}} and \ch{Mn^{IV}}: the ccp \ch{VCl4} structure lies only 21 meV/atom (0.105 eV/f.u.) above the \ch{VCl4} cluster, and the O1-like \ch{MnCl4} structure lies only 13 meV/atom (0.065 eV/f.u.) above the Cl-dimerized structure. These results anticipate that highly delithiated chlorides in which the transition-metal cations reach a high oxidation state will likely not be particularly stable, and may be susceptible to \ch{Cl-} oxidation and subsequent dimerization.


\FloatBarrier
\begin{figure*}[h]
\centering
\includegraphics[width=\linewidth]{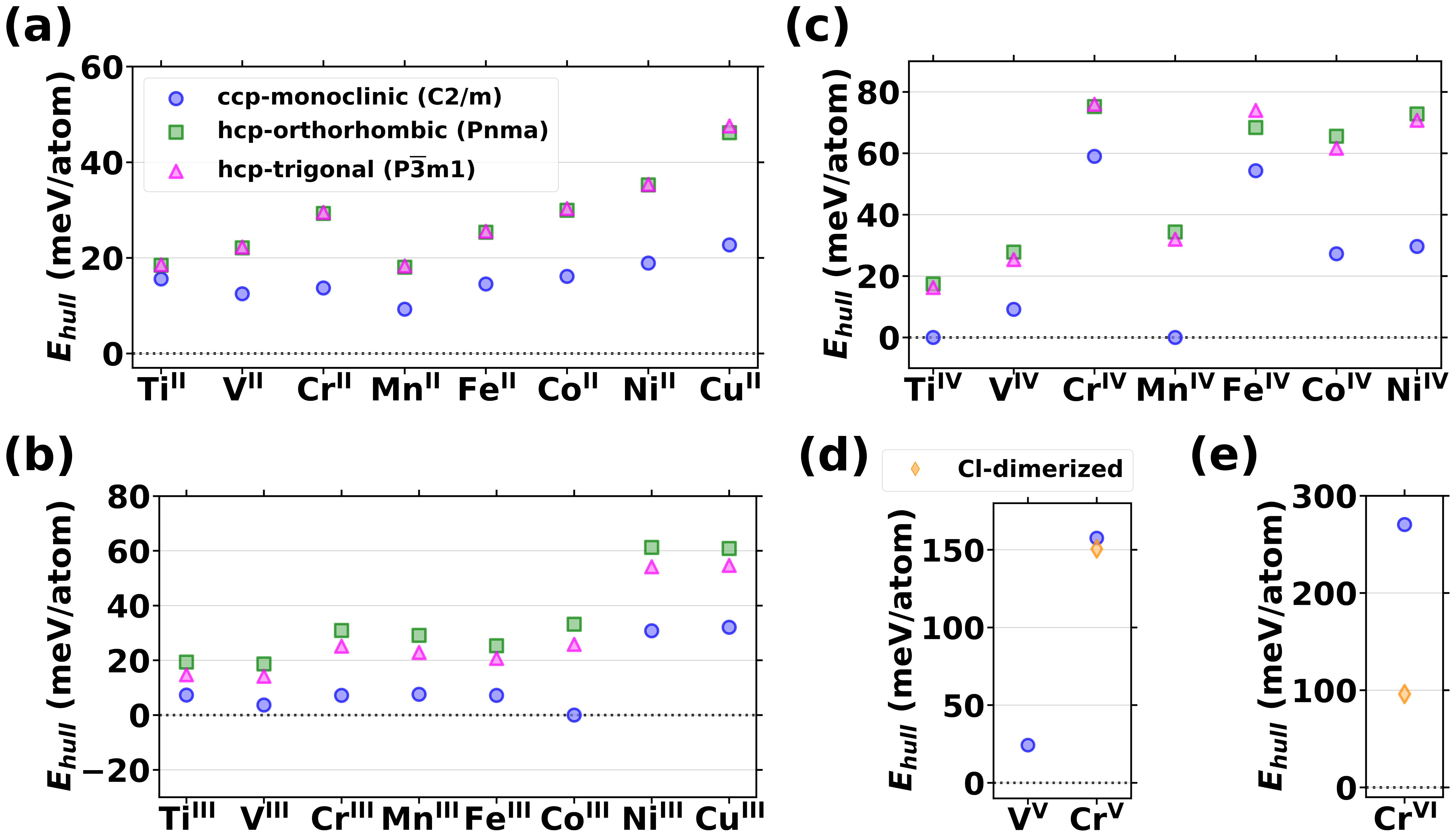}
\caption{$E_{\mathrm{hull}}$ of (a) \ch{Li4M^{II}Cl6}, (b) \ch{Li3M^{III}Cl6}, (c) \ch{Li2M^{IV}Cl6}, (d) \ch{LiM^{V}Cl6}, and (e) \ch{M^{VI}Cl6} across different polymorphs and 3\textit{d} transition-metal species.}
\label{fig:LixMCl6_Ehull} 
\end{figure*}

\FloatBarrier
\begin{figure*}[t]
\centering
\includegraphics[width=0.7\linewidth]{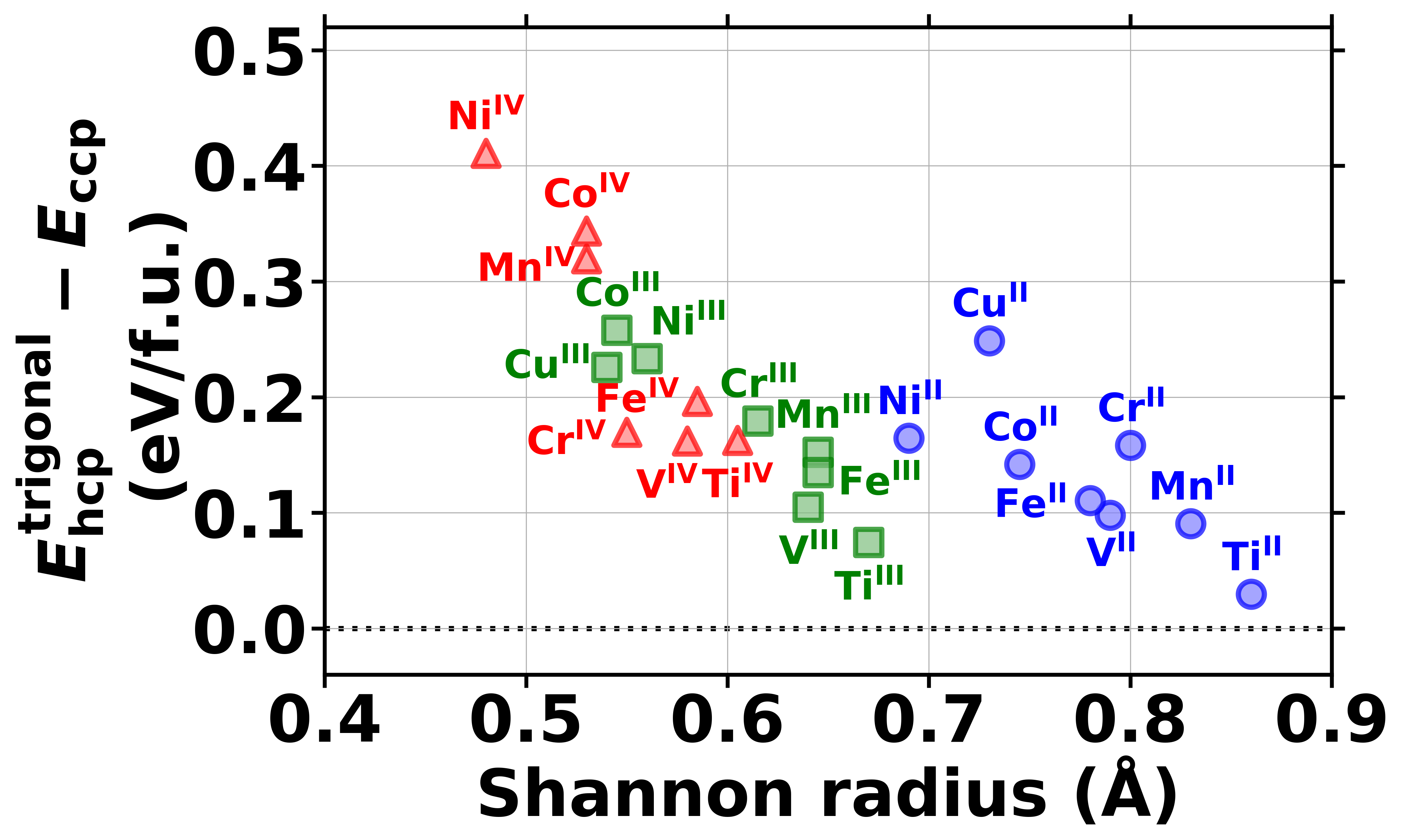}
\caption{Relative phase stability of \ch{Li_{4-x}MCl6} between ccp and hcp-trigonal structures across the transition-metal species and their formal oxidation states.}
\label{fig:LixMCl6_Shannon_radii_deltaE_trigonal}
\end{figure*}

Figure \ref{fig:LixMCl6_Ehull} shows the $E_{\mathrm{hull}}$ of \ch{Li_{4-x}MCl_6} compounds in either the ccp-monoclinic (C2/m), hcp-orthorhombic (Pnma), or hcp-trigonal (P$\overline{3}$m1) structure. Figures \ref{fig:LixMCl6_Ehull} (a)-(c) show $E_{\mathrm{hull}}$ values from II to IV formal oxidation states. Our results show that the ccp-monoclinic phase has the lowest energy across the formal oxidation states from II to IV. In Figures \ref{fig:LixMCl6_Ehull} (d) and (e), we also consider the higher oxidation states up to VI for V and Cr in the ccp-monoclinic framework because it exhibits a lower energy than the hcp frameworks at the lower oxidation states. Our results in Figures \ref{fig:LixMCl6_Ehull} (d) and (e) show that Cr is quite unstable at oxidation states above IV: $E_{\mathrm{hull}}$ values for topotactically delithiated ccp structures are over 150 meV/atom and 270 meV/atom for \ch{LiCr^{V}Cl6} and \ch{Cr^{VI}Cl6}, respectively. Similar to the \ch{M^{IV}Cl4} structures (see Figure \ref{fig:LixMCl4_Ehull} (c)), we find that Cl--Cl dimerization is possible at these highly charged \ch{LiCr^{V}Cl6} and \ch{Cr^{VI}Cl6} structures (the latter is schematically shown in Figure \ref{fig:Cl_dimerization} (c)). As shown in Figure \ref{fig:LixMCl6_Ehull} (e), the Cl--Cl dimerization reduces the energy by 174 meV/atom for \ch{Cr^{VI}Cl6} relative to the topotactically delithiated ccp structure.

Figure \ref{fig:LixMCl6_Shannon_radii_deltaE_trigonal} shows the energy difference between the ccp-monoclinic and hcp-trigonal phases as a function of cation size. For a given oxidation state, the relative phase stability exhibits an approximately linearly decreasing trend with increasing Shannon radius, indicating that the hcp Cl anion packing becomes more competitive with the ccp packing for a larger cation. Our result aligns with previous findings in redox-inactive halides, which have shown that the relative stability of ccp and hcp structures is correlated with the ionic potential (the ratio of cation charge to cation radius) \cite{Li2024_ccp_hcp_ionic_potential,Wang2024_ionic_potential}. In general, 3\textit{d} transition metals have smaller ionic radii than the redox-inactive metals typically used in solid-state electrolytes. For example, the largest 3\textit{d} cations, \ch{Ti^{3+}} (0.67 Å) and \ch{Ti^{4+}} (0.605 Å), are still smaller than redox-inactive cations in the corresponding charge state, such as \ch{In^{3+}} (0.80 Å), \ch{Y^{3+}} (0.90 Å), \ch{Sc^{3+}} (0.745 Å), and \ch{Zr^{4+}} (0.72 Å). As a result, all eREAL materials investigated in this study exhibit a preference for ccp over hcp anion packing.

\begin{figure*}[t]
\centering
\includegraphics[width=\linewidth]{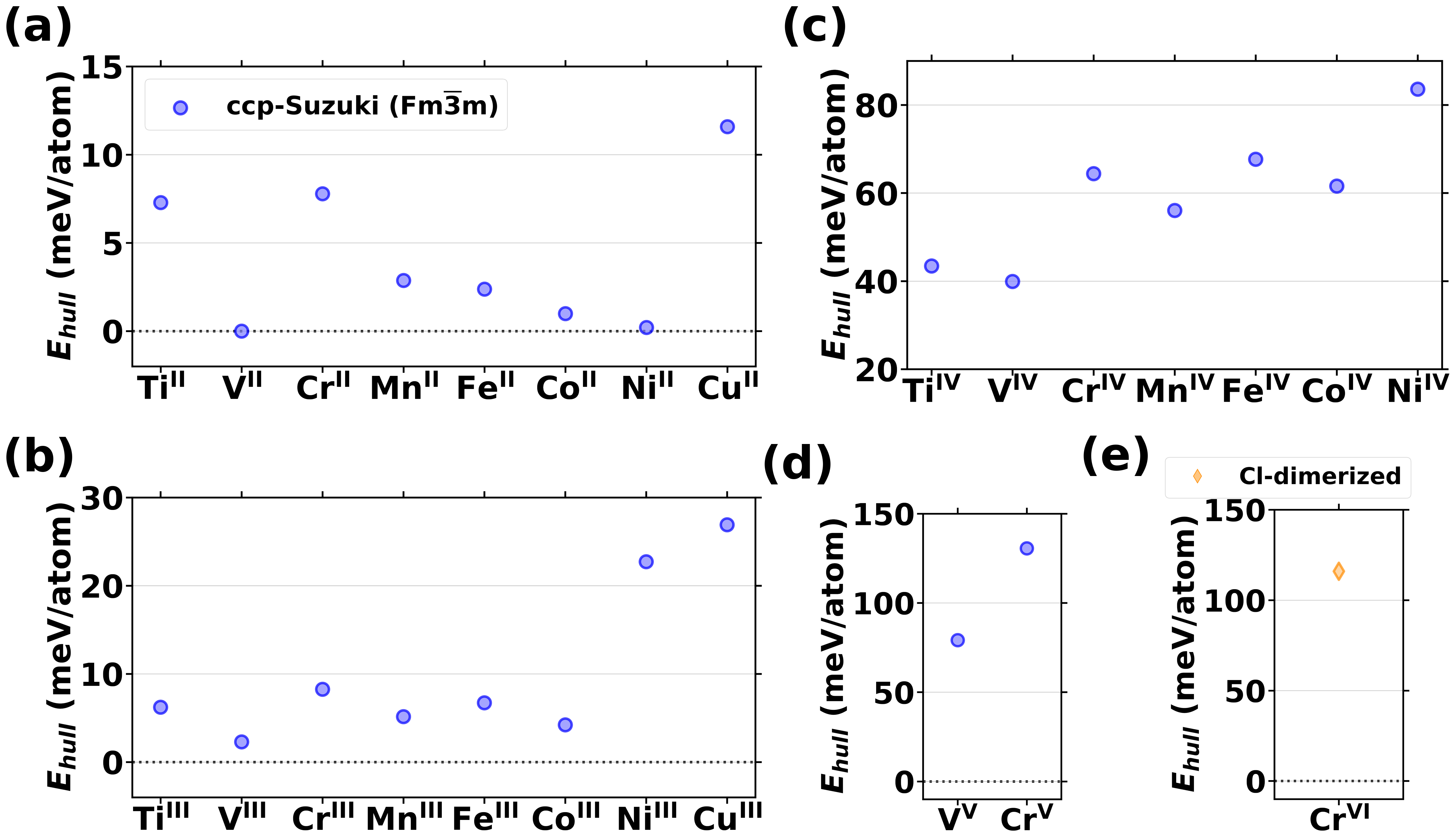}
\caption{$E_{\mathrm{hull}}$ of Suzuki-type (a) \ch{Li6M^{II}Cl8}, (b) \ch{Li5M^{III}Cl8}, (c) \ch{Li4M^{IV}Cl8}, (d) \ch{Li3M^{V}Cl8}, and (e) \ch{Li2M^{VI}Cl8} structures with different 3\textit{d} transition-metal cations.}
\label{fig:LixMCl8_Ehull} 
\end{figure*}

Figure \ref{fig:LixMCl8_Ehull} shows $E_{\mathrm{hull}}$ of Suzuki-type \ch{Li_{6-x}MCl_8} structures at formal oxidation states from II to VI. These structures share the same symmetry as rocksalts (Fm$\overline{3}$m). Similar to the other two composition families (i.e., \ch{Li_{2-x}MCl4} and \ch{Li_{4-x}MCl6}), the $E_{\mathrm{hull}}$ value of \ch{Li_{6-x}MCl8} increases with oxidation, especially at the IV formal oxidation state and above. As illustrated in Figure \ref{fig:Cl_dimerization} (d), Cl anions in \ch{Li2Cl^{VI}Cl8} spontaneously dimerize upon relaxation. 

Overall, our results suggest that topotactic delithiation is more accessible at lower (II-III) oxidation states but becomes thermodynamically unfavorable at the IV formal oxidation state and above, where Cl anions tend to dimerize, particularly for structures with later transition metals. The Cl--Cl dimerization is a sign of the onset of Cl anion redox and potential \ch{Cl2} gas release at high voltages, which sets an upper limit on accessible cation redox potential. This is similar to oxygen redox activity in oxide cathodes at high voltages \cite{Ni2006_O2_release,Seo2016_O_redox,Luo2016_O_redox,Assat2018_anion_redox,House2021_O_redox_review}. The electronic structure reason for the easy dimerization in chlorides and the difference with oxides will be discussed in the next section.

\FloatBarrier
\subsection{Cation and anion redox activities in eREAL materials}
\FloatBarrier
\begin{figure*}[t]
\centering
\includegraphics[width=0.8\linewidth]{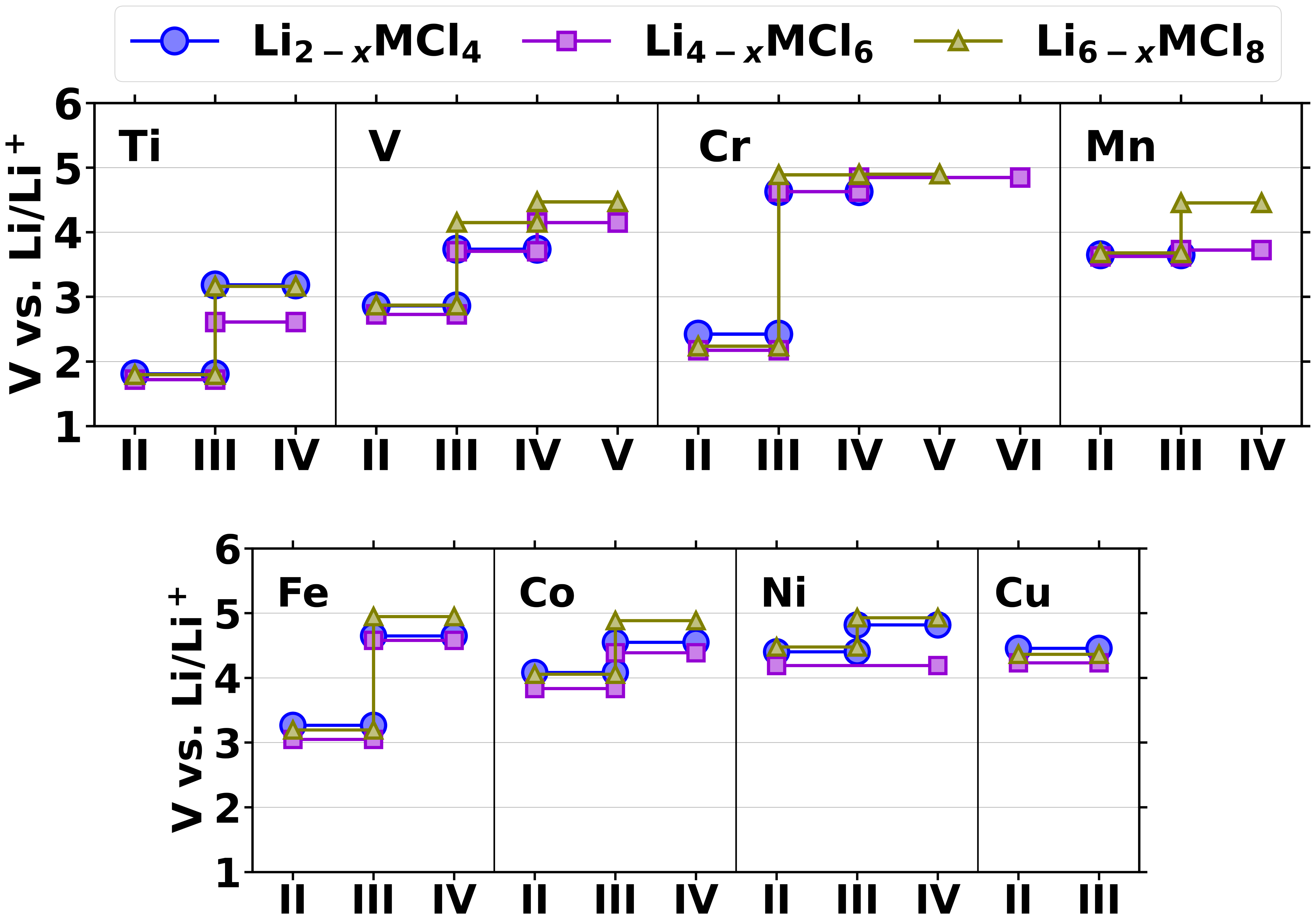}
\caption{\textbf{Topotactic average intercalation voltages in Li--M--Cl chemical systems} The average voltages are shown as averages between oxidation states of the corresponding redox couples. The blue, purple, and yellow colors represent the average voltages for \ch{Li_{2-x}MCl4}, \ch{Li_{4-x}MCl6}, and \ch{Li_{6-x}MCl8}, respectively, where M stands for redox-active transition metals.}
\label{fig:average_voltages} 
\end{figure*}

To investigate the potential redox activity of eREAL materials, we calculate the average Li intercalation voltages of eREAL materials with 3\textit{d} transition metals by taking the energy difference between the lithiated and delithiated structures, following the well-established computational method \cite{Aydinol1997_voltage_calculation_method,Urban2016_voltage_method}. 
We evaluate the average voltages in two ways. In the first approach, we only allow topotactic delithiation within ccp anion lattices (referred to as the topotactic voltage), so that we investigate the effect of the transition-metal species and the metal-to-Cl ratio. In the second approach, we allow the large distortions and Cl--Cl dimerization that are shown in the previous section to break the symmetry for high oxidation states. Lifting the constraint to maintain the ccp framework lowers the energy at high oxidation states, leading to voltages (referred to as non-topotactic voltages) that are lower than the topotactic ones but may better represent the practical values in experiments.

Figure \ref{fig:average_voltages} shows the topotactic average voltages between successive oxidation states for each 3\textit{d} transition metal, evaluated for structures with three different metal-to-Cl ratios. The topotactic voltage shows an increasing trend as one moves from early to late transition metals, which is expected from the 3\textit{d} orbital energy lowering at higher atomic numbers \cite{Aydinol1997_voltage_calculation_method}. Notably, our results indicate that the redox potential is largely insensitive to the metal-to-Cl ratio, which only leads to different arrangements of Li ions and transition-metal cations within the same face-centered cubic (FCC) chlorine framework for most of the compositions. This finding suggests that the cation redox potential in these phases is primarily governed by the [\ch{MCl6}] octahedral ligand field and the local Li site energy. This conclusion is further supported by our calculations shown in Figure SX in the Supporting Information, which indicates that altering the anion packing from ccp to hcp for \ch{Li_{4-x}MCl6} yields a difference of less than 0.1 V in average voltage for most redox couples. Figure \ref{fig:average_voltages} reveals some exceptions to this structure-insensitive behavior. For \ch{Li_{4-x}NiCl6}, we predict a single plateau at 4.2 V between the \ch{Ni^{II}} and \ch{Ni^{IV}} oxidation states, with the intermediate \ch{Ni^{III}} state being unstable, unlike the stepped voltage curves for \ch{Li_{2-x}NiCl4} and \ch{Li_{6-x}NiCl8} which have well defined \ch{Ni^{III}} states. The II/IV double redox potential in \ch{Li_{4-x}NiCl6} is $\sim$0.6--0.7 V lower than the III/IV voltages in the \ch{Li_{2-x}NiCl4} and \ch{Li_{6-x}NiCl8} composition series. For \ch{Li_{4-x}MnCl6}, the \ch{Mn^{III}} state is stable, but the average voltages of \ch{Mn^{II}}/\ch{Mn^{III}} and \ch{Mn^{III}}/\ch{Mn^{IV}} redox couples are very similar at $\sim$3.7 V, which is $\sim$0.8 V lower than the III/IV voltage in \ch{Li_{6-x}MnCl8}. For \ch{Li_{4-x}TiCl6}, the \ch{Ti^{III}}/\ch{Ti^{IV}} plateau is clearly separated from the \ch{Ti^{II}}/\ch{Ti^{III}} plateau but also lies more than $\sim$0.5 V lower than in the \ch{Li_{2-x}TiCl4} and \ch{Li_{6-x}TiCl8} composition series. Together, these results indicate that the highest oxidation states (i.e., \ch{Ti^{IV}}, \ch{Mn^{IV}}, and \ch{Ni^{IV}}) are stabilized more in the \ch{Li_{4-x}MCl6} structural framework than in the other two types of structures with different metal-to-Cl ratios, although the origin of this high stability is unclear.

\FloatBarrier
\begin{figure*}[t]
\centering
\includegraphics[width=\linewidth]{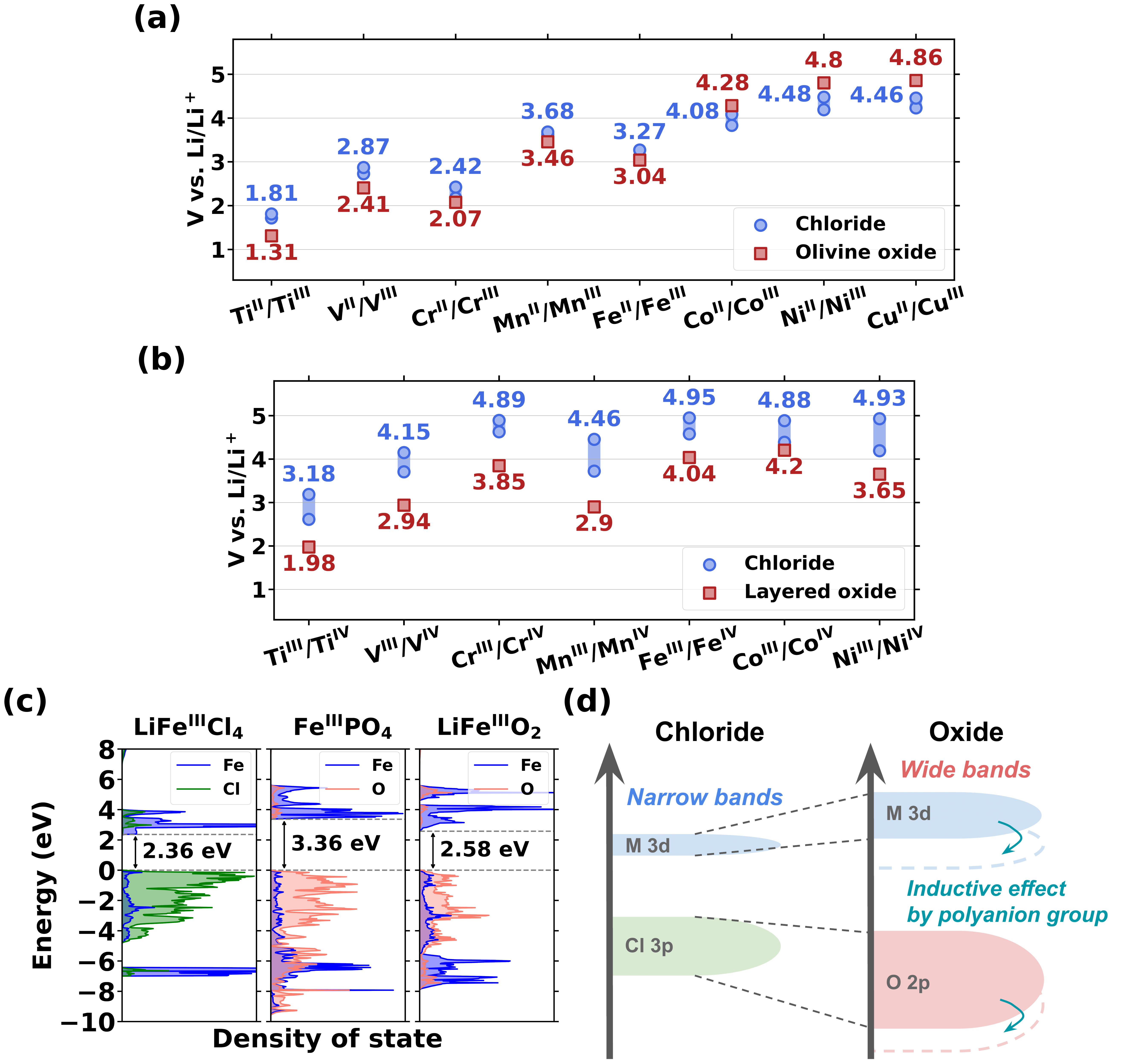}
\caption{\textbf{Comparison of computed redox potentials in chloride and oxide systems} (a) Topotactic average voltage of 
\ch{M^{II}/M^{III}} redox couples in the chloride and olivine (\ch{Li_{1-x}MPO4}) structures. (b) Topotactic average voltage of 
\ch{M^{III}/M^{IV}} redox couples in the chloride and layered oxide (\ch{Li_{1-x}MO2}) structures. The voltages for chlorides are shown as a range across the \ch{Li_{2-x}MCl4}, \ch{Li_{4-x}MCl6}, and \ch{Li_{6-x}MCl8} structures, and the highest voltages are labeled. (c) Projected DOS of \ch{LiFe^{III}Cl4}, \ch{Fe^{III}PO4}, and \ch{LiFe^{III}O2}. The zero of energy is placed at the valence band maximum. The DOS is calculated using the HSE06 hybrid functional. (d) Schematic of metal \textit{d}-anion \textit{p} orbital hybridization and inductive effect.}
\label{fig:chloride_oxide_comparison} 
\end{figure*}

The computed topotactic average voltages shown in Figure \ref{fig:average_voltages} for some transition-metal species are underestimated compared to the few known experimental values, which may be attributed to the residual self-interaction error inherent in the r$^2$SCAN functional \cite{Isaacs2020_SCAN_benchmark}. As has been previously argued, self-interaction error in the \textit{d}-states artificially increases the energy of transition-metal cations with more electrons and hence lowers the voltages of oxidation \cite{Zhou2004_self_interaction_error}. For example, the experimentally measured \ch{Fe^{II/III}} and \ch{Mn^{II/III}} redox potentials are $\sim3.7$ and $\sim4.2$ V, respectively \cite{Liu2024_Li2FeCl4_exp,Tanibata2025_redox_exp}, which are 0.4--0.5 eV higher than our calculation. To gain a qualitative understanding of the redox potentials in eREAL systems, we compare in Figure \ref{fig:chloride_oxide_comparison} the computed topotactic average voltages for chlorides with similarly computed values for typical oxide cathode materials, i.e., olivine \ch{Li_{1-x}MPO4} and layered oxides \ch{Li_{1-x}MO2}. The voltages for chlorides are shown as a range across the three metal-to-anion ratios. As shown in Figure \ref{fig:chloride_oxide_comparison} (a), the \ch{Li_{2-x}MCl4} eREAL materials with transition metals earlier than Co exhibit a II/III average voltage that is even higher than that in the olivine structure. The phosphate group in olivine structures is known to increase the cation redox potentials compared to those of pure oxides due to the inductive effect of the strongly covalent P--O bonds, which lowers the covalency of M--O bonds \cite{Manthiram1989_inductive_effect, Hautier2011_phosphate}. Low covalency with an anion keeps the energy of the metal-3\textit{d} levels low, hence the voltage high \cite{Yang2021_cathode_review}. Our results suggest that the bond ionicity of M--Cl bonds are comparable to or even higher than the M--O(--P) bonds. Figure \ref{fig:chloride_oxide_comparison} (b) compares the III/IV redox potentials of the eREALs and layered oxides. As expected, the layered oxides have significantly lower average voltages than chlorides for the equivalent oxidation state, as the M--O bonds in layered oxides are much more covalent. 


The ionic nature of M--Cl bonds can be assessed from electronic structure calculations. Figure \ref{fig:chloride_oxide_comparison} (c) compares the projected density of states (pDOS) of \ch{LiFeCl4}, \ch{FePO4}, and \ch{LiFeO2} at the III oxidation state, as calculated using the HSE06 hybrid functional to achieve higher accuracy. For all these compounds, the valence band maximum is composed predominantly of anion \textit{p} orbital character (Cl 3\textit{p} or O 2\textit{p}), whereas the conduction band minimum features mainly Fe 3\textit{d} orbitals. The degree of covalency in the metal–anion bonds can be inferred from the electronic bandwidth: the more covalent M–O bonds in the oxides give rise to more delocalized states, resulting in wider bands compared with those in the chloride. Despite its more localized states, \ch{LiFeCl4} also exhibits a smaller band gap, which can also be attributed to its stronger bond ionicity which brings the Fe 3\textit{d} band closer in energy to the Fermi level. The effect of \textit{d}--\textit{p} orbital hybridization and the inductive effect by the phosphate group on the electronic band structure are illustrated in Figure \ref{fig:chloride_oxide_comparison} (d).

\FloatBarrier
\begin{figure*}[h]
\centering
\includegraphics[width=0.8\linewidth]{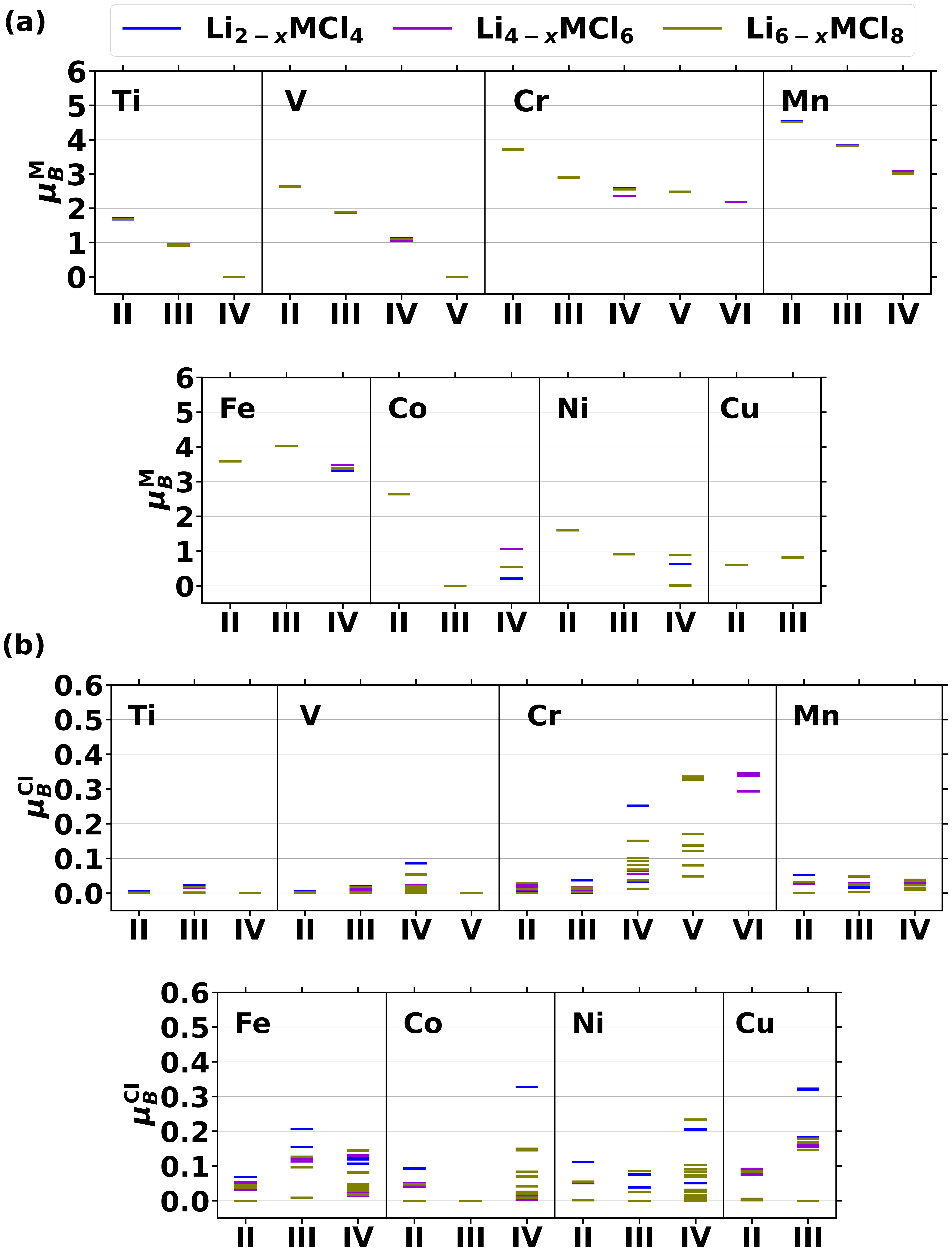}
\caption{Atomic magnetic moments of (a) transition-metal cations and (b) Cl anions in topotactically delithiated structures at different formal oxidation states. The blue, purple, and yellow colors represent the magnetic moments for \ch{Li_{2-x}MCl4}, \ch{Li_{4-x}MCl6}, and \ch{Li_{6-x}MCl8}, respectively, where M stands for 3\textit{d} transition metals.}
\label{fig:Li--M--Cl_magmom} 
\end{figure*}

In principle, the high cation redox potential makes eREAL materials promising for achieving high energy density. However, the application of eREAL materials in composite cathodes is likely to be hindered by the activation of Cl anion redox and \ch{Cl2} gas release above $\sim4$ V, which precludes access to redox couples involving later transition metals (such as Co, Ni, and Cu) or those at higher oxidation states (such as III/IV) \cite{Tanibata2025_redox_exp,Liu2024_Li2FeCl4_exp}. To confirm the activation of Cl oxidation, we calculate in Figure \ref{fig:Li--M--Cl_magmom} the atomic magnetic moments on the transition-metal cations and Cl anions in the structures considered in the topotactic average voltage calculations in Figure \ref{fig:average_voltages}. As has been argued previously, the use of magnetic moments on transition metals is a better gauge of their valence state than integrated charge density \cite{Reed2002_magmom}. Figure \ref{fig:Li--M--Cl_magmom} (a) shows that the magnetic moments on transition-metal cations generally decrease with increasing oxidation, consistent with progressive electron extraction upon delithiation. Reasonable exceptions from this decreasing trend are observed for high-spin \ch{Fe^{III}}, which has a higher magnetic moment than \ch{Fe^{II}}, and for low-spin \ch{Co^{III}}, which has a zero magnetic moment, lower than that of \ch{Co^{IV}}. In addition, \textit{d}$^9$ \ch{Cu^{II}} has a lower magnetic moment than \textit{d}$^8$ \ch{Cu^{III}}. These magnetic moments are nonetheless consistent with the expected spin-polarized electronic configuration of the respective cations \cite{Jia2022_persona}. At low oxidation states (e.g., II or III), the magnetic moment on transition-metal cations shown in Figure \ref{fig:Li--M--Cl_magmom} (a) reflect charge states that are largely consistent with the formal oxidation states. Evidence for Cl oxidation is observed at higher formal oxidation states (e.g., IV, V, or VI), where the charge states of some transition-metal cations suggested by these magnetic moments deviate significantly from the assigned formal oxidation states. For example, Figure \ref{fig:Li--M--Cl_magmom} (a) shows that nominally \ch{Cr^{VI}} cations in the \ch{CrCl6} structures retain non-zero magnetic moments, although \ch{Cr^{VI}} cations should be \textit{d}$^0$ and thus non-magnetic. This suggests that these \ch{Cr^{VI}} cations are at lower charge states and that the charge compensation has to be occurring on Cl anions. Figure \ref{fig:Li--M--Cl_magmom} (a) also shows similar deviations between the formal oxidation states and the true cation charge states for \ch{Co^{IV}} and \ch{Ni^{IV}}. The low-spin \ch{Co^{4+}} and \ch{Ni^{4+}} are expected to have magnetic moments of 1 and 0 $\mu_{\mathrm{B}}$, respectively, while \ch{Co^{4+}} in the high-spin state should have a magnetic moment of 3 $\mu_{\mathrm{B}}$. However, the predicted magnetic moments for some \ch{Co^{IV}} cations fall below 1 $\mu_{\mathrm{B}}$ and for some \ch{Ni^{IV}} cations exceed 0 $\mu_{\mathrm{B}}$, suggesting that the true charge states of these IV-oxidation-state cations are lower than 4+. In addition to the cation magnetic moments, we can also infer Cl anion oxidation directly from the Cl magnetic moments. As shown in Figure \ref{fig:Li--M--Cl_magmom} (b), nonzero magnetic moments are predicted on Cl anions for structures with later transition metals or at higher formal oxidation states. The small but nonzero Cl magnetic moments at low oxidation states, such as those for \ch{Fe^{II}}, \ch{Mn^{II}}, \ch{Co^{II}}, \ch{Ni^{II}}, and \ch{Cu^{II}} are due to charge transfer from Cl to the cation through the weak hybridization and not evidence for Cl oxidation. At higher oxidation states, the Cl magnetic moment clearly increases, indicating the oxidation of Cl anions. The Cl oxidation in the topotactically delithiated structures also explain why Cl--Cl dimerization is energetically favorable at the highly charged states as discussed in Figures \ref{fig:LixMCl4_Ehull}, \ref{fig:LixMCl6_Ehull}, and \ref{fig:LixMCl8_Ehull}. 

\FloatBarrier
\begin{figure*}[t]
\centering
\includegraphics[width=0.8\linewidth]{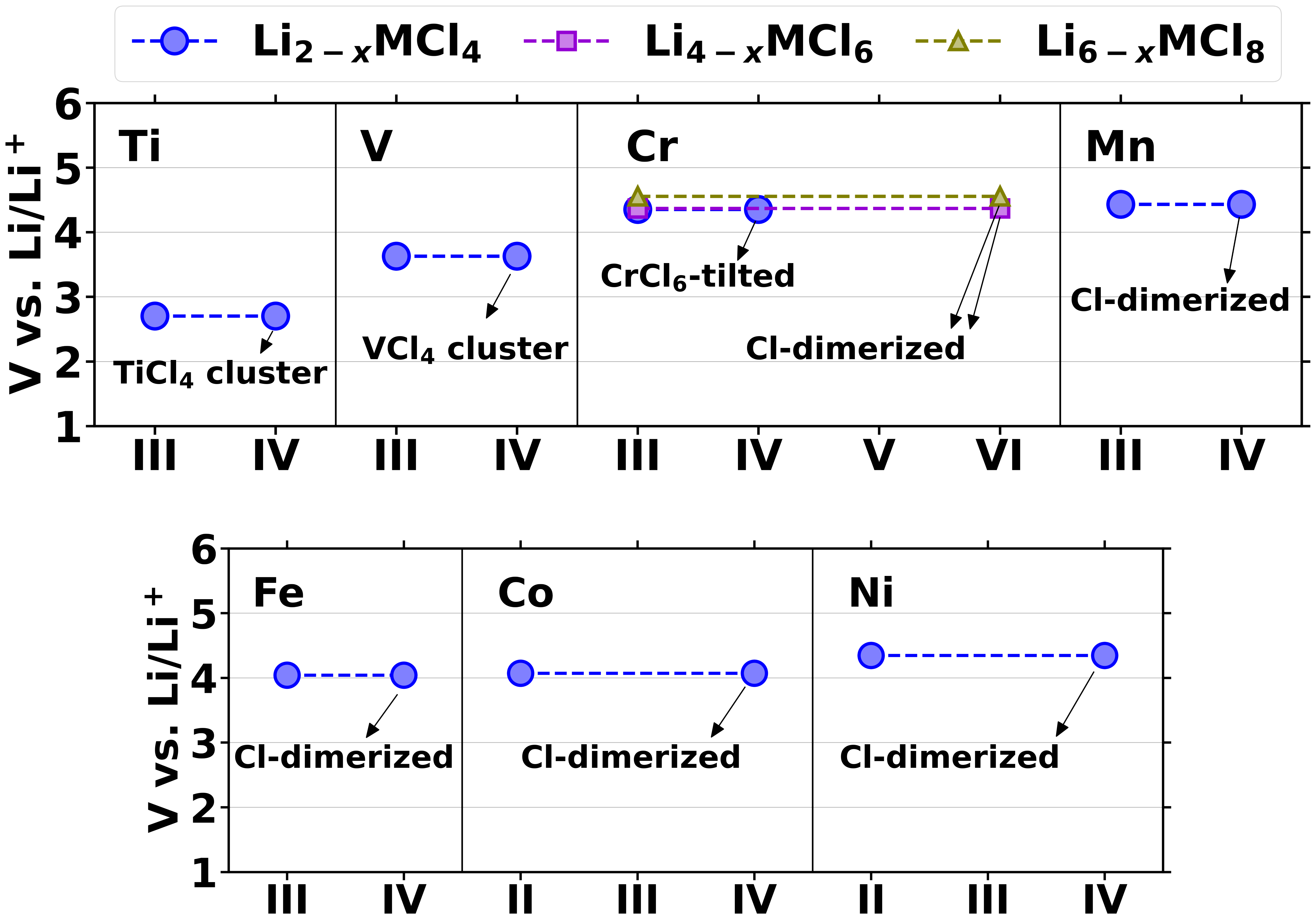}
\caption{\textbf{Non-topotactic average intercalation voltages in Li--M--Cl chemical systems} The average voltages are shown as averages between oxidation states of the corresponding redox couples. The blue, purple, and yellow colors represent the average voltages for \ch{Li_{2-x}MCl4}, \ch{Li_{4-x}MCl6}, and \ch{Li_{6-x}MCl8}, respectively, where M stands for redox-active transition metals. The types of structural distortion at the charged states are labeled. }
\label{fig:average_voltages_nontopo} 
\end{figure*}

In Figure \ref{fig:average_voltages_nontopo}, we evaluate the non-topotactic average voltages of the redox couples for which severe structural distortions are predicted at highly charged states. Our results in Figure \ref{fig:LixMCl4_Ehull}(c) show that the structures at IV formal oxidation state break the close-packed symmetry through either the formation of [\ch{MCl4}] clusters, tilting of [\ch{MCl6}] octahedra, or Cl--Cl dimerization. Cl--Cl dimerization is also observed for \ch{Cr^{V}} and \ch{Cr^{VI}} states, as shown in Figures \ref{fig:LixMCl6_Ehull} (d) and (e) and Figure \ref{fig:LixMCl8_Ehull} (e). Figure \ref{fig:average_voltages_nontopo} shows that for the III/IV redox couples in the \ch{Li_{2-x}MCl4} structures, the structural distortions reduce the average voltage by 0.48 V, 0.1 V, 0.28 V, and 0.6 V for Ti, V, Cr, and Fe, respectively, relative to the topotactic values shown in Figure \ref{fig:average_voltages} . As shown in Figure \ref{fig:average_voltages_nontopo}, we predict that the Cl--Cl dimerization occurs at a voltage as low as 4.0 V (for Fe and Co), which is in good agreement with previous experimental observations that \ch{Cl2} gas release occurs when halide-based cathodes are charged above $\sim$4 V \cite{Tanibata2025_redox_exp,Liu2024_Li2FeCl4_exp}. Figure \ref{fig:average_voltages_nontopo} shows that Cl--Cl dimerization leads to a double redox process between the II and IV oxidation states for Co and Ni in the \ch{Li_{2-x}MCl4} structures, with the average voltages 0.48 V and 0.47 V lower than the corresponding topotactic III/IV voltages for Co and N, respectively (see Figure \ref{fig:average_voltages}). One can also think of this as a disproportionation of the \ch{M^{3+}} state (i.e., III formal oxidation state) into \ch{M^{2+}} (i.e., II formal oxidation state) and \ch{Cl^0} (i.e., IV formal oxidation state), which is typically referred to as a ligand hole in the transition-metal oxide literature \cite{Zaanen1985_ligand_hole}. Figure \ref{fig:average_voltages_nontopo} also shows that in the \ch{Li_{4-x}CrCl6} and \ch{Li_{6-x}CrCl8} structures, Cl--Cl dimerization results in a triple redox between the III and VI oxidation states at 4.4 V and 4.6 V, respectively, which are both lower than the topotactic voltage of 4.8 V obtained for the Cr IV/VI redox couple in the \ch{Li_{4-x}CrCl6} structures (see Figure \ref{fig:average_voltages}).

\begin{figure*}[h]
\centering
\includegraphics[width=\linewidth]{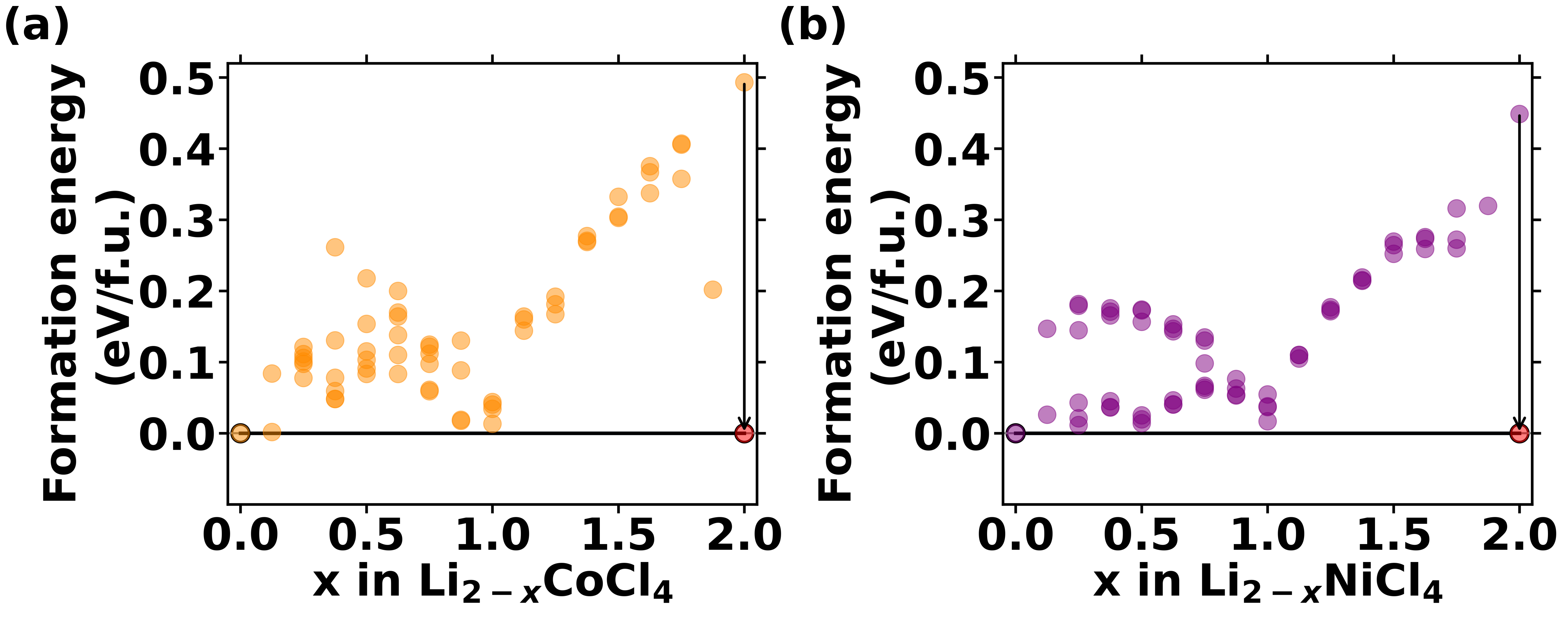}
\caption{Formation energies of (a) \ch{Li_{2-x}CoCl4} and (b) \ch{Li_{2-x}NiCl4} with respect to the most stable endpoint structures (at x $=0$ and x $=2$), where Li ions are extracted from a fully lithiated supercell containing 16 Li in steps of 0.125 Li per formula unit. The fully delithiated (a) \ch{CoCl4} and (b) \ch{NiCl4} structures (i.e., x $=2$) lower their energies by Cl--Cl dimerization, which are highlighted with red circles on the hulls. The dashed lines are the topotactic convex hulls among the undimerized structures. }
\label{fig:Li2CoCl4_Li2NiCl4_voltage_profiles} 
\end{figure*}

In Figure \ref{fig:average_voltages}, we have shown a weak dependence of the topotactic cation redox potentials on the structural and compositional factors (e.g., metal-to-Cl ratio, cation ordering, and anion packing). This is consistent with the very flat voltage profiles often observed for eREAL materials in experiments \cite{Liu2024_Li2FeCl4_exp, Fu2025_all_in_one, Wang2023_LTC, Song2024_LVC}. To probe this behavior, we selectively calculate voltage profiles for \ch{Li_{2-x}CoCl4} and \ch{Li_{2-x}NiCl4} by removing one Li at a time from a supercell containing 16 Li (i.e., in steps of 0.125 Li per f.u.) and enumerating Li/vacancy arrangements at each composition. At the fully delithiated state (x $= 2$), we consider both the topotactically delithiated ccp structure and the Cl-dimerized structure, with the latter predicted to be the lowest-energy structure for both \ch{CoCl4} and \ch{NiCl4}, as shown in Figure \ref{fig:LixMCl4_Ehull} (c). Figures \ref{fig:Li2CoCl4_Li2NiCl4_voltage_profiles} (a) and (b) show the computed formation energies of \ch{Li_{2-x}CoCl4} and \ch{Li_{2-x}NiCl4} structures, where the energy of intermediate compounds are referenced to the most stable endpoint structures at x $= 0$ and x $= 2$. As can be seen, only the endpoint structures appear on the hull for both \ch{Li_{2-x}CoCl4} and \ch{Li_{2-x}NiCl4} composition series. As a result of Cl--Cl dimerization, the voltage profiles of these structures appear as flat plateaus between II and IV formal oxidation states, which are exactly the same as the non-topotactic average voltage profiles shown in Figure \ref{fig:average_voltages_nontopo}. Even between the \ch{Co^{II}} (\ch{Ni^{II}}) and \ch{Co^{III}} (\ch{Ni^{III}}) undimerized state, no intermediate compositions appear stable, indicating that the voltage profile would remain flat between these integer valence states even in the absence of Cl–Cl dimerization. It is remarkable that no states with mixed valance are on the convex hull, which is unlike most layered oxide cathodes that go through multiple Li/vacancy ordered phases upon delithiation \cite{Aydinol1998_LCO, Dompablo2003_LNO,Ma2011_NaMnO2,Yang2021_cathode_review}. The chlorides seem to behave more like \ch{Li_{1-x}FePO4} in which the strong attraction between \ch{Li+} and its accompanying reduced transition metal form essentially a bound neutral carrier, which removes the repulsive \ch{Li+}-\ch{Li+} interaction that causes the intermediate ordered states in layered oxides \cite{Malik2013_LFP}. Our results indicate that the delithiation of \ch{Li_{2-x}CoCl4} and  \ch{Li_{2-x}NiCl4} beyond the II oxidation state likely induces severe structural degradation, making reversible cycling of these compounds unlikely. The flat voltage plateaus also have an implication for the mixed ionic-electronic conduction behavior: when an eREAL catholyte is paired with an active material, these two redox-active materials are likely to contribute separately to the capacity at distinct voltage values or ranges, making it challenging to maintain the mixed-valence state that is crucial for electronic conductivity, without some more compositional engineering.

\FloatBarrier
\begin{figure*}[t]
\centering
\includegraphics[width=0.7\linewidth]{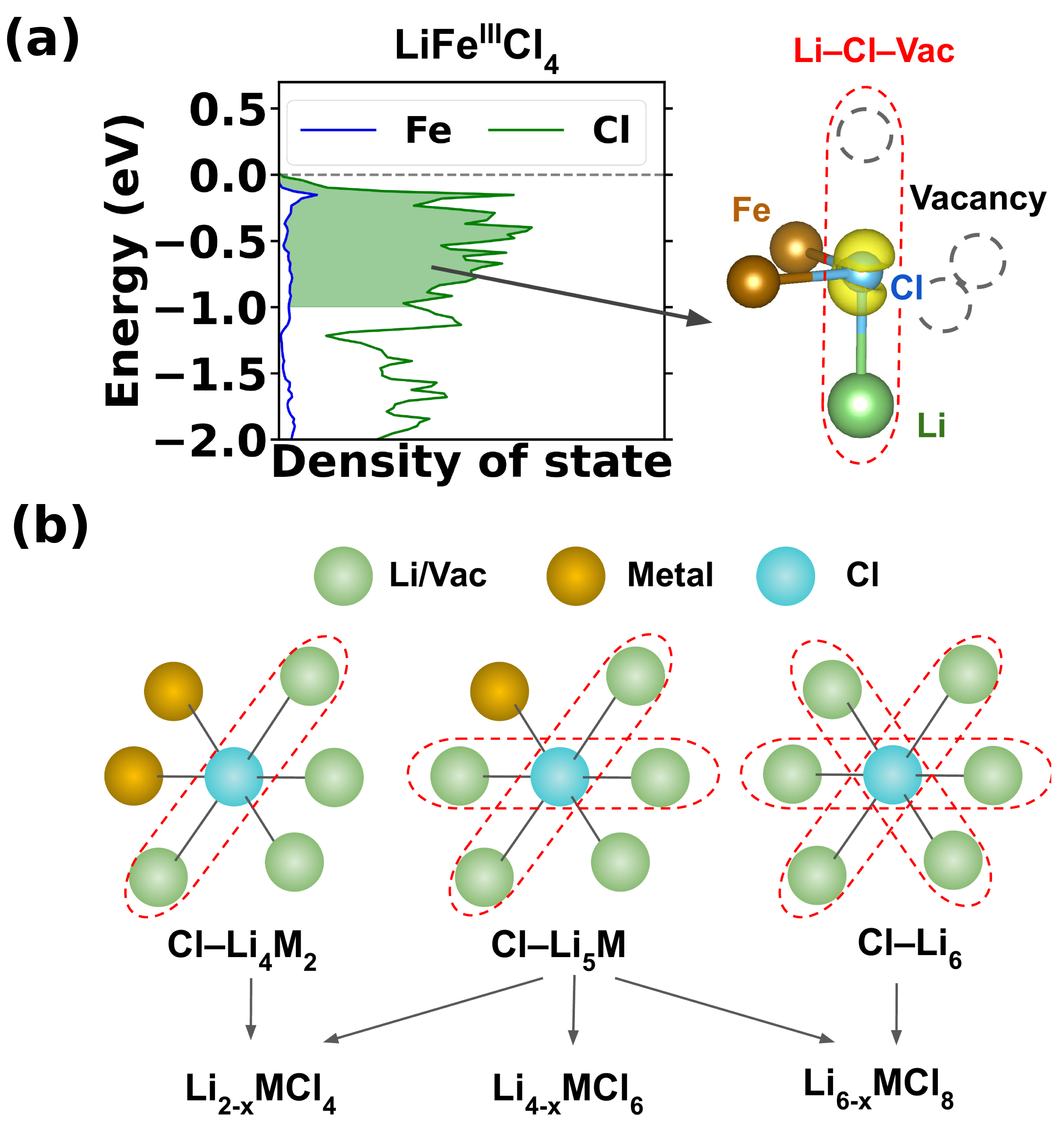}
\caption{\textbf{Orphaned Cl 3\textit{p} orbitals in close-packed halide compounds} (a) Projected DOS of \ch{LiFe^{III}Cl4} and the isosurface of the charge density (yellow) around the Cl anion coordinated by one Li, three vacancies, and two Fe, in the energy range of 0 to -1 eV. (b) Three types of local Cl environments exist in the \ch{Li_{2-x}MCl4}, \ch{Li_{4-x}MCl6}, and \ch{Li_{6-x}MCl8} structures with ccp anion lattice. The local environments are denoted as Cl--\ch{Li4M2}, Cl--\ch{Li5M}, or Cl--\ch{Li6}, depending on the number of coordinated transition metals. The linear Li/Vac--Cl--Li/Vac motif is highlighted with red dashed ellipsoids, where "Vac" stands for a vacancy.}
\label{fig:Li-Cl-Li} 
\end{figure*}

\FloatBarrier
Several factors contribute to the facile activation of the Cl anion redox. In a solid, the anion level is typically the bonding state of the metal-\textit{d} and anion-\textit{p} orbital overlap. In oxides, the strong hybridization of the O anion with the metal pushes its states down in energy, thereby protecting them from oxidation to higher voltages than those at which the \ch{O2} molecule evolves from metal-free lithium oxides such as \ch{Li2O} \cite{Yang2021_cathode_review}. Since the M--Cl bond is more ionic, the \ch{Cl-} state is less protected and will oxidize closer to the molecular energy level. The low metal-to-Cl ratio also plays a role as there are simply less metals present to protect the \ch{Cl-} orbital. Figure \ref{fig:Li-Cl-Li}(a) shows that the Cl 3\textit{p} states at the valence band maximum in \ch{LiFe^{III}Cl4} correspond to electron charge densities that resemble orphaned Cl 3\textit{p} orbitals along the linear Li--Cl--Vacancy motifs. Due to poor overlap with Li or vacancy sites, Cl 3\textit{p} states have little oxidation protection from hybridization as is the case for O 2\textit{p} states in Li-excess oxides \cite{Seo2016_O_redox,Luo2016_O_redox}. Figure \ref{fig:Li-Cl-Li}(b) shows the different types of Cl anion local environments present in the chlorides with varying metal-to-Cl ratios. As the metal-to-Cl ratio decreases from 1:4 in \ch{Li_{2-x}MCl4} to 1:8 in \ch{Li_{6-x}MCl8}, less metals surround the Cl anions, leading to more Li/Vac–Cl–Li/Vac motifs in the chloride structure. In \ch{Li_{2-x}MCl4}, there are two types of Cl local environments coordinated either by two transition metals and four Li (referred to as Cl–Li$_4$M$_2$) or by one transition metal and five Li (referred to as Cl–Li$_5$M). In comparison, the Cl local environments in \ch{Li_{6-x}MCl8} structures are either Cl–Li$_5$M or fully coordinated with Li (referred to as Cl–Li$_6$). \ch{Li_{4-x}MCl6} structures lie in between these two extremes, with only the Cl–Li$_5$M local environment present. The increased number of Li ions (or vacancies) surrounding the central Cl anion further weakens the hybridization and promotes Cl anion redox. In comparison, in \ch{LiMO2} layered oxides, the double negatively charged \ch{O^{2-}} allows the incorporation of more transition metals, so that all oxygen anions are bonded to at least three transition metals. This atomic arrangement enhances the covalent interactions between transition-metal cations and O anions, moving the anion oxidation up to much higher voltage unless cation disorder occurs, Li-excess is incorporated, or transition-metal migration creates vacancies in the transition-metal layer \cite{Seo2016_O_redox,Luo2016_O_redox,Gent2017_TM_migration}. Finally, it has been shown that the lack of transition metal bonding makes it easier for the oxidized anion to dimerize, thereby further lowering the oxidation potential \cite{House2021_O_redox_review, Assat2018_anion_redox}. It has also been argued that once the dimerization occurs, oxidation/reduction becomes highly irreversible or occurs with large hysteresis \cite{Ni2006_O2_release, Assat2017_hysteresis,House2020_hesteresis,House2020_hysteresis_first_cycle}.

\begin{figure*}[h]
\centering
\includegraphics[width=\linewidth]{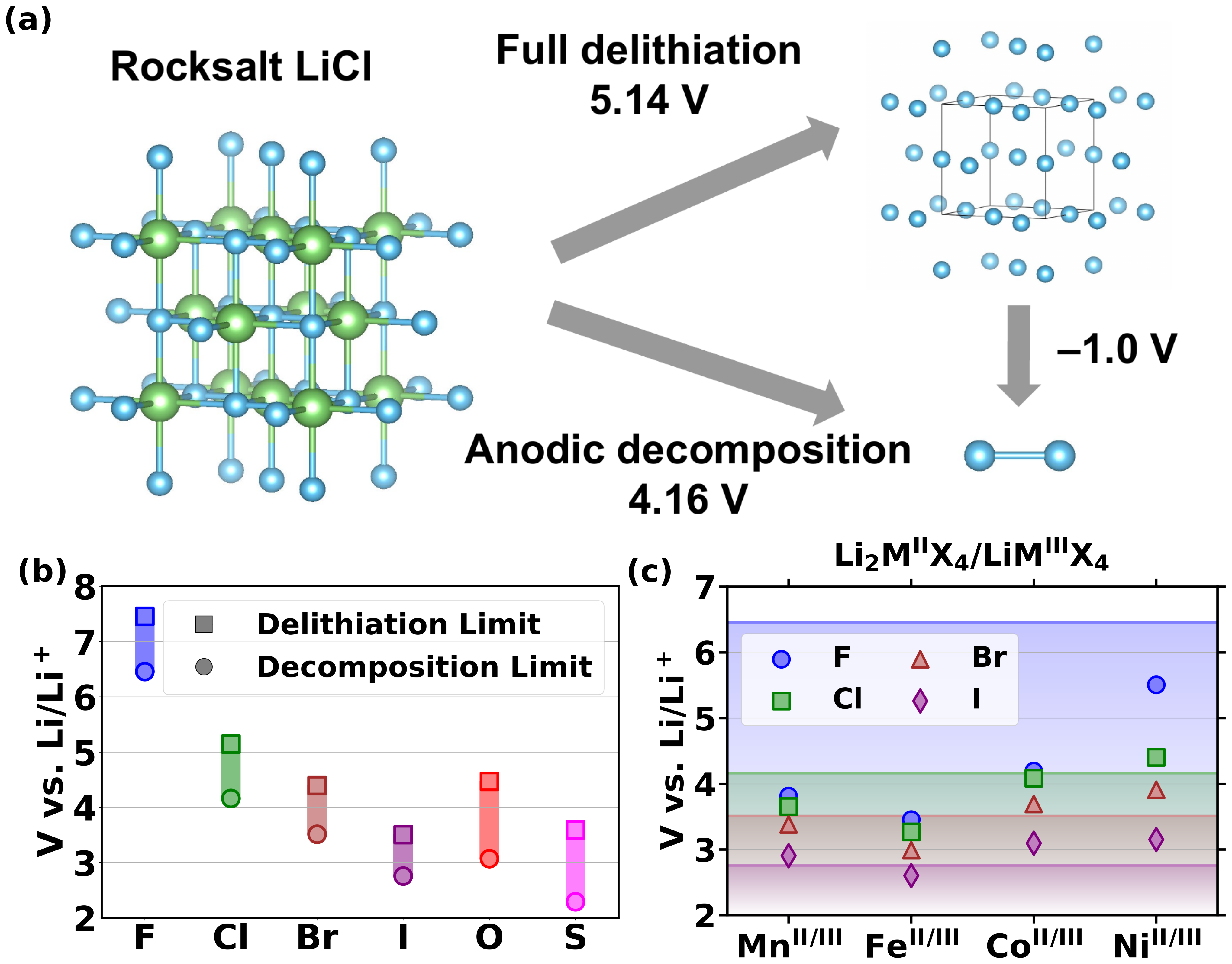}
\caption{\textbf{Anion oxidation limits} (a) Theoretical oxidation limits of Cl anion are estimated by either topotactic delithiation of rocksalt LiCl to a fixed Cl FCC sublattice (delithiation limit) or anodic decomposition to bulk Li metal and \ch{Cl2} gas molecules in vacuum (decomposition limit). (b) Comparison of delithiation and decomposition oxidation limits for LiX (X = F, Cl, Br, I) and \ch{Li2Y} (Y = O, S). Decomposition limits for \ch{LiF}, \ch{LiCl}, and \ch{Li2O} are referenced to the DFT energy of bulk Li metal and isolated molecules \ch{F2}, \ch{Cl2}, and \ch{O2} in vacuum, respectively. Decomposition limits for \ch{LiBr}, \ch{LiI}, and \ch{Li2S} are referenced to the DFT energy of bulk Li metal and fully relaxed condensed-phase molecular unit cells containing multiple \ch{Br2}, \ch{I2}, and \ch{S8} molecules, respectively. (c) Effect of halogen anion substitution on the average intercalation voltage of \ch{M^{II/III}} redox couples in the \ch{Li_{2-x}MCl_4} composition, which are given as data points. The voltage range below the decomposition limits of rocksalt LiX is highlighted a color-coded background with a gradient (i.e., blue, green, brown, and purple for LiF, LiCl, LiBr, and LiI, respectively).}
\label{fig:anion_redox} 
\end{figure*}

To estimate the bare anion redox potential, we perform the following computational "experiments" (see Figure \ref{fig:anion_redox}(a)). In the first scenario, starting from rocksalt LiCl, we fully delithiate without relaxing the remaining FCC Cl anion lattice, which by its nature cannot have transition-metal redox. This oxidation voltage (termed the delithiation limit) is calculated as $V_{\mathrm{deli}}=-(E_{\mathrm{LiCl}}-E_{\mathrm{Li}}-E_{\mathrm{Cl}}^{\mathrm{FCC}})/F$, where $E_{\mathrm{LiCl}}$, $E_{\mathrm{Li}}$, and $E_{\mathrm{Cl}}^{\mathrm{FCC}}$ are the DFT energy (per f.u.) of rocksalt LiCl, Li metal, and the Cl FCC sublattice, and $F$ is the Faraday constant. Figure \ref{fig:anion_redox}(a) shows that this topotactic delithiation which oxidizes \ch{Cl-} to \ch{Cl^0} occurs at 5.14 V. This value represents an absolute upper voltage limit for Cl anion redox when the Cl orbitals have no stabilization at all from hybridization. In the second scenario, Cl anions are allowed to dimerize and form \ch{Cl2} molecules in vacuum. This oxidation limit (termed the decomposition limit) is calculated as $V_{\mathrm{decomp}}=-(E_{\mathrm{LiCl}}-E_{\mathrm{Li}}-0.5E_{\mathrm{Cl_2}})/F$, where $E_{\mathrm{Cl_2}}$ is the DFT energy of an isolated \ch{Cl2} molecule in vacuum. As Figure \ref{fig:anion_redox}(a) shows, we find a value of 4.16 V for this process, indicating that the oxidation voltage drops by $\sim$1 V when unprotected \ch{Cl-} states are oxidized and form \ch{Cl2} gas molecules.

Equivalent anion oxidation limits of other lithium halides and lithium chalcogenides, i.e., LiX (X = F, Br, I) and \ch{Li2Y} (Y = O, S), are also calculated, and shown in Figure \ref{fig:anion_redox}(b). The topotactic delithiation limit of these structures is calculated using the same procedure as for LiCl, i.e., by fixing the FCC anion lattice upon delithiation. The decomposition limit for \ch{LiF} or \ch{Li2O} is referenced to the DFT energy of bulk Li metal and an isolated \ch{F2} or \ch{O2} molecule in vacuum, respectively. For \ch{LiBr}, \ch{LiI}, or \ch{Li2S}, the decomposition limit is referenced to Li metal plus a condensed-phase molecular unit cells containing multiple \ch{Br2}, \ch{I2}, and \ch{S8} molecules, respectively, because these molecules are liquids or solids at room temperature. These molecular unit cells are obtained from the Materials Project \cite{Jain2013_MP,Horton2025_MP}, with lattice parameters and atomic positions fully relaxed using the computational settings in this work. As shown in Figure \ref{fig:anion_redox}(b), within both the halide and chalcogenide families, the oxidation limits decrease moving down the periodic table, consistent with the decreasing electronegativity of anion species. LiCl has higher delithiation (5.14 V) and decomposition (4.16 V) limits than those of \ch{Li2O}, which are 4.47 and 3.07 V, respectively. The DFT-predicted delithiation limit (i.e., the upper limit of oxidation stability) of \ch{Li2O} is very close to the experimentally measured voltage plateau at $\sim4.5$ V during the first charge of Li-rich layered oxide cathodes, which is interpreted as an O anion oxidation process without \ch{O2} gas evolution \cite{Luo2016_O_anion_redox_JACS}. Our result provides an important insight: the fact that oxide cathodes can operate at higher voltages than chloride cathodes is not a consequence of a higher intrinsic oxidation limit. Instead, the superior oxidation stability of oxides stems from the covalent bonding with multiple metals which lowers the oxygen level, and combined with the steric limitations imposed by the layered topology which prevents neighboring O anions from easily coming close to each other and forming O--O dimers. In other words, while the oxidation stability of eREAL materials is limited by dimerization of oxidized Cl, layered oxides can undergo topotactic delithiation to high voltages without triggering irreversible \ch{O2} gas evolution.

It is known that substituting a less electronegative anion can enhance the covalency of metal-anion bonds, and thus lower the cation redox potential. Combined with our findings above, this leads to multiple potential strategies that might mitigate the anion oxidation issue and extend the reversible cation redox capacity in halides. The first strategy is substituting Cl with less electronegative halogens, such as Br and I, such that the cation redox potentials are systematically lowered. However, Br and I anions also exhibit lower anion oxidation limits than Cl, and thus the Br and I anion redox may still compete with the cation redox. The second strategy is substituting Cl with the more electronegative F, such that the decomposition limit that leads to anion dimerization is increased, although this also raises the cation redox potentials. To test these hypotheses, we calculate the intercalation voltages for \ch{Li2M^{II}X4}/\ch{LiM^{III}X4} redox couples with combinations of selected transition metals (M = Mn, Fe, Co, or Ni) and four halogens (X = F, Cl, Br, or I). Figure \ref{fig:anion_redox}(c) shows the average intercalation voltage for each cation redox in structures with different anion species (shown as markers). Voltage ranges below the anodic decomposition limits ($V_{\mathrm{decomp}}$) of rocksalt halides LiX (X = F, Cl, Br, or I) are highlighted with colored backgrounds. In this figure, if a cation redox potential lies below an anodic decomposition limit, then the cation redox can likely be accessed during delithiation without triggering the anion redox-induced structural decomposition; conversely, if it lies above this limit, anion oxidation is expected before the onset of cation redox. Our results in Figure \ref{fig:anion_redox}(c) indicate that in Br- and I-based compositions, both cation and anion redox potentials decrease by similar amounts relative to those of the corresponding chlorides. For the \ch{Co^{II/III}} and \ch{Ni^{II/III}} redox couples, the cation redox potentials remain higher than the $V_{\mathrm{decomp}}$ of \ch{LiBr} and \ch{LiI}, respectively, making them unlikely to be accessible. In comparison, the predicted \ch{Fe^{II/III}} redox potential lies 0.53 V and 0.15 V below the $V_{\mathrm{decomp}}$ of LiBr and LiI, respectively. However, these margins are smaller than that of the chlorides (0.89 V). We note that since $V_{\mathrm{decomp}}$ of lithium halides does not involve transition metal orbitals, it is better represented by r$^2$SCAN functional \cite{Kothakonda2022_SCAN_r2SCAN_benchmark,Zhang2018_SCAN_benchmark}, whereas r$^2$SCAN underestimates the \ch{Fe^{II/III}} redox voltage in chlorides by $\sim$0.4 V \cite{Liu2024_Li2FeCl4_exp,Tanibata2025_redox_exp,Isaacs2020_SCAN_benchmark}. Taking this underestimation into consideration, our results suggest that \ch{Fe^{II/III}} cation redox is more accessible in chlorides (as experimentally verified \cite{Liu2024_Li2FeCl4_exp,Fu2025_all_in_one}), possibly in bromides, but unlikely in iodides. Figure \ref{fig:anion_redox}(c) further shows that the \ch{Mn^{II/III}} cation redox potential is even higher than that of \ch{Fe^{II/III}}, making the accessibility of \ch{Mn^{II/III}} cation redox in bromides and iodides even more limited. Therefore, our results indicate that substituting Cl with Br or I is unlikely to provide strong protection against anion oxidation at voltages where cation redox is active. In contrast, Figure \ref{fig:anion_redox}(c) shows that F substitution raises the cation redox potentials by a much smaller amount than it raises the anodic decomposition limit. The cation redox potentials increase by only 0.16 V, 0.19 V, and 0.11 V for the \ch{Mn^{II/III}}, \ch{Fe^{II/III}}, and \ch{Co^{II/III}} redox couples, respectively, relative to the chlorides, while the $V_{\mathrm{decomp}}$ rises to above 6.45 V. The average \ch{Ni^{II/III}} voltage increases the most upon F substitution (by 1.1 V), however, it remains considerably below the F--F dimerization voltage. Therefore, our results suggest that from the perspective of mitigating anion redox, F substitution stands out as a more effective strategy than other anion substitutions in eREAL materials. Nonetheless, full substitution of Cl by F is likely to lead to very poor Li-ion conductivity, but partial substitution may dilute Cl anions enough to prevent substantial amount of Cl--Cl dimerization to occur.

\FloatBarrier
\begin{figure*}[t]
\centering
\includegraphics[width=0.8\linewidth]{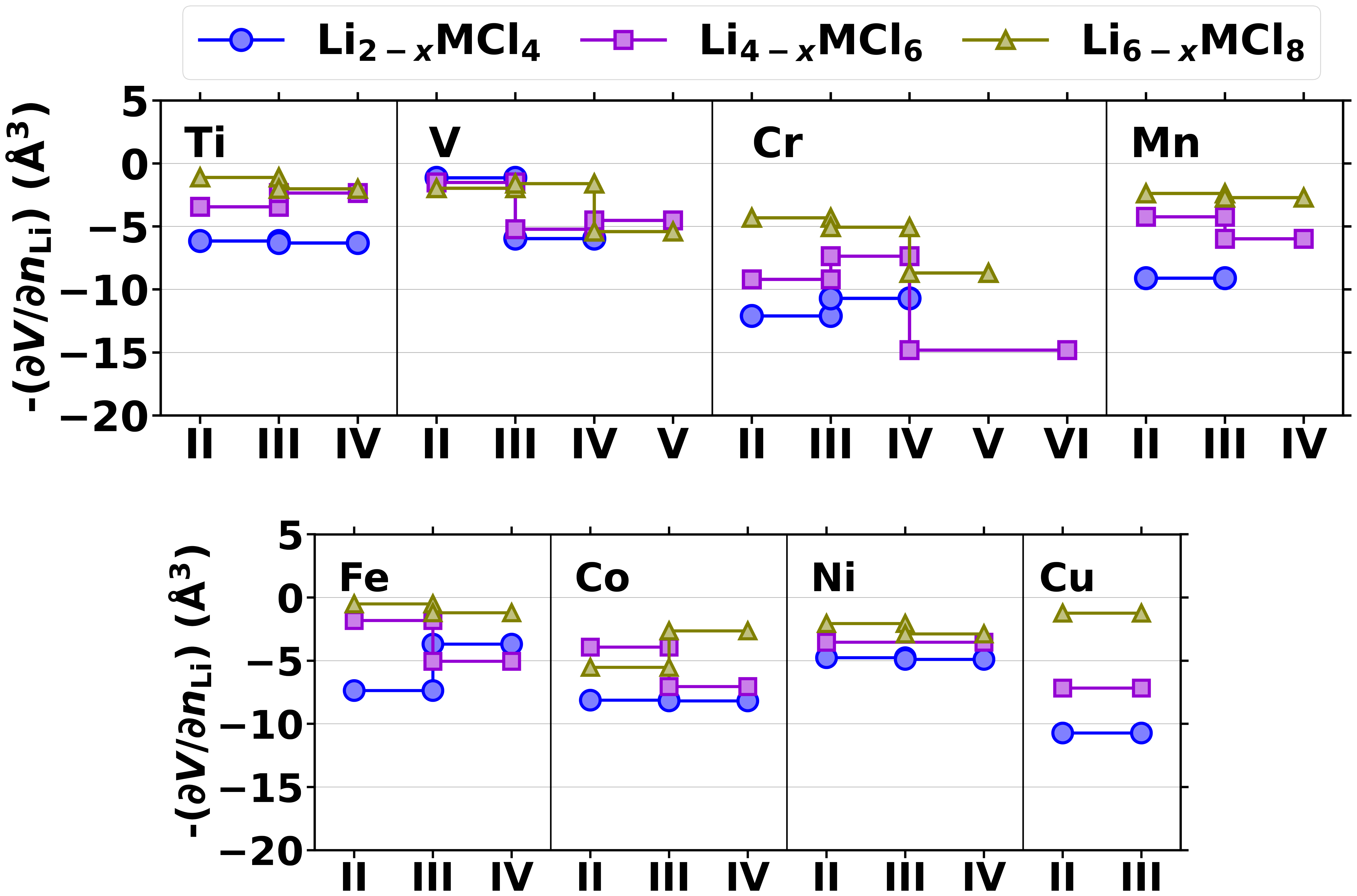}
\caption{\textbf{Partial molar volumes of Li upon topotactic delithiation.} The partial molar volumes are shown as averages between oxidation states of the corresponding redox couples. The blue, purple, and yellow colors represent the partial molar volumes for \ch{Li_{2-x}MCl4}, \ch{Li_{4-x}MCl6}, and \ch{Li_{6-x}MCl8}, respectively, where M stands for 3\textit{d} transition metals. }
\label{fig:Li-M-Cl_partial_molar_volumes} 
\end{figure*}

Mechanical degradation during cycling is another common issue in solid-state batteries \cite{Ko2026_review_practical_SSE}. Due to the high stiffness and low plasticity of solid components, maintaining intimate electrode-electrolyte interfacial contact is more difficult than in liquid-electrolyte cells, which reduces the effective contact area for Li ion and electron transport. In addition, volume change during cycling exerts additional stress that may induce crack formation, which further degrades the interfacial contact. To estimate the volume change of eREAL materials upon cycling, we calculate the partial molar volumes for each redox couple using the method introduced in Ref\cite{Zhao2022_zero_strain}, where the partial molar volume for a given redox couple is defined as the change in crystal structure volume per intercalated Li. Figure \ref{fig:Li-M-Cl_partial_molar_volumes} shows the resulting partial molar volume profiles across 3\textit{d} transition-metal species for the redox couples identified from the topotactic average voltage calculations shown in Figure \ref{fig:average_voltages}. The y-axis displays the negative partial molar volume ($-(\partial V/\partial n_{\mathrm{Li}})$), where a more negative value indicates greater structural contraction upon delithiation. As can be seen, a higher metal-to-Cl ratio generally results in a larger expansion/contraction in volume, i.e., $-(\partial V/\partial n_{\mathrm{Li}})$ becomes more negative, upon Li extraction. Among the II/III redox couples, \ch{Cr^{II}/Cr^{III}} in \ch{Li_{2-x}CrCl4} exhibits the most negative $-(\partial V/\partial n_{\mathrm{Li}})$ value of $-12$ Å$^3$, which corresponds to a volume contraction of 8.5 $\%$. This may be rationalized by considering that the partial molar volume is determined by both the reduction in the cation radius when it is oxidized and the ratio of Li removed to Li remaining in the structure. A lower metal-to-Cl ratio corresponds to a higher LiCl content per formula unit, which is a redox-inactive component in the structure. The higher fraction of LiCl dilutes the concentration of redox-active transition metals and vacancies created during Li extraction, thereby buffering the mechanical strain during electrochemical cycling. Therefore, a lower metal-to-Cl ratio might offer greater resistance to mechanical degradation during cycling. This is in contrast to the negligible effect of metal-to-Cl ratio on the electrochemical redox potential, as shown in Figure \ref{fig:average_voltages}. An exception is observed in Figure \ref{fig:average_voltages} for the Cr IV-VI redox couple, which exhibits the most negative $-(\partial V/\partial n_{\mathrm{Li}})$ value among all redox couples. This is because the fully delithiated \ch{CrCl6} structure consists of isolated [\ch{CrCl6}] octahedra that contract the lattice through van der Waals interactions. However, as discussed above, the topotactically delithiated \ch{CrCl6} structure is highly unstable and likely releases \ch{Cl2} gas upon charging, in which case the $-(\partial V/\partial n_{\mathrm{Li}})$ value cannot be well defined.

\FloatBarrier
\section{Discussion}

Using electrochemically active solid-state conductors is an attractive option to significantly increase the energy density of composite cathodes. Chlorides are of particular interest due to their high Li-ion conductivity (in some materials) and relatively high metal redox potentials. In this work, we investigate in detail which structures are likely to form in Li--M--Cl systems and what controls the potential of the redox process in chlorides. We demonstrate that most eREAL compositions containing first-row 3\textit{d} transition metals are likely to adopt a ccp anion framework, consistent with the smaller ionic radius of most 3d transition metals as compared to main group elements. In agreement with prior findings \cite{Li2024_ccp_hcp_ionic_potential,Wang2024_ionic_potential}, we find that the energy difference between hcp and ccp anion-packing structures systematically decreases as the transition-metal ionic radius increases. The stability of ccp packing may benefit the Li transport, since ccp frameworks generally exhibit higher Li-ion conductivity than hcp frameworks in halide solid-state conductors \cite{Li2024_ccp_hcp_ionic_potential}.

Within the ccp-type \ch{Li2MCl4} structures, Li ions occupy either octahedral or tetrahedral sites, leading to two different polymorphs, i.e., orthorhombic (Cmmm) and inverse spinel (Imma). We show that the energetically favored polymorph depends on the transition metal, but the energy difference between the two polymorphs is small across the 3\textit{d} transition-metal species ($<$ 20 meV/atom). We also find that as the material becomes cation deficient, e.g., by delithiation, there is a tendency for all remaining ions to occupy octahedral sites. These findings agree with the experimental observation that in a Li-deficient structure \ch{Li_{1.3}Fe_{1.2}Cl4}, all Li and Fe occupy octahedral sites as characterized by a combination of X-ray diffraction, neutron diffraction, and pair distribution function techniques \cite{Fu2025_all_in_one}.

The only \ch{LiMCl4} structures in which we find a tetrahedrally coordinated transition metal is \ch{LiFeCl4}, whose lowest-energy structure adopts an hcp-monoclinic polymorph containing tetrahedrally coordinated \ch{Fe^{III}}. However, the energy difference between the ccp and hcp-monoclinic \ch{LiFeCl4} phases is relatively small (22 meV/atom), which aligns with the general understanding that high-spin \textit{d}$^5$ \ch{Fe^{III}} has a low octahedral-site stabilization energy, and therefore exhibits a weaker preference for a particular coordination environment than many other transition-metal cations \cite{Yang2021_cathode_review, Manthiram2025_cation_redox_review}. Consistent with this, reversible Fe migration from octahedral to tetrahedral sites has been observed experimentally when Fe in the ccp \ch{Li_{1.6-x}Fe_{1.2}Cl4} is charged from II to III formal oxidation states \cite{Fu2025_all_in_one}.

Our results also expose potential issues when eREAL compounds are combined with cathodes that may require high-voltage charging. Upon further delithiation toward \ch{MCl4}, we find that the topotactically delithiated ccp structures become highly unstable and exhibit a substantial driving force for structural distortions (including the formation of [\ch{MCl4}] clusters, tilting of [\ch{MCl6}] octahedra, and Cl--Cl dimerization). More broadly, we find that irrespective of the metal-to-Cl ratio, the thermodynamic stability of eREAL phases decreases substantially (i.e., $E_{\mathrm{hull}}$ increases) upon oxidation, particularly at formal oxidation states of IV and above. Among these distortions, Cl--Cl dimerization has the largest driving force for late transition metals at highly charged states and likely leads to structural decomposition with \ch{Cl2} gas evolution \cite{Tanibata2025_redox_exp,Liu2024_Li2FeCl4_exp}.

These stability issues for charged eREAL compounds and their limits on the accessible charging voltage arise from the highly ionic nature of the metal--Cl bond and the low metal-to-Cl ratio, which together render the \ch{Cl-} anion more vulnerable to oxidation and dimerization than in equivalent oxides. We demonstrate that the cation redox potential in chlorides is generally higher than that in oxides, including the polyanion \textit{p}hosphate group-based compounds such as olivines. The higher cation redox potential suggests that the strongly ionic metal--Cl bond keeps the metal anti-bonding state at low energy. While the high cation redox potentials of eREAL materials makes them promising for high-voltage applications, our results also show that these materials will be susceptible to anion redox-induced degradation above $\sim4.2$ V (i.e., the r$^2$SCAN-predicted oxidation limit for Cl--Cl dimerization), which will limit the reversible capacity of eREAL cathodes. Importantly, our calculations reveal that the intrinsic oxidation voltage limits (i.e., both the topotactic oxidation of Cl anion or the oxidation when Cl--Cl dimerization occurs in the structure) of unhybridized \ch{Cl^-} anions is actually higher than that of similarly unhybridized \ch{O^{2-}} anions. Nonetheless, \ch{Cl2} release in chlorides at $\sim4$ V is widely observed in experiments \cite{Tanibata2025_redox_exp,Liu2024_Li2FeCl4_exp}, whereas oxides can usually sustain voltages up to $\sim 4.5$ V without \ch{O2} evolution even in those Li-rich compounds with non-bonding O 2\textit{p} states \cite{Luo2016_O_anion_redox_JACS,Seo2016_O_redox,Luo2016_O_redox,House2021_O_redox_review}. We attribute the lower practical stability of chlorides against oxidation to three factors. First, the ionic metal--Cl bonding sets the bonding state (with a dominant Cl anion character) relatively high in energy. Second, the single negative charge on \ch{Cl^-} anions leads to a low metal-to-anion ratio than in oxides, limiting the hybridization necessary for stabilizing Cl 3\textit{p} orbitals. The linear Li/Vac–Cl–Li/Vac motifs are abundant in close-packed chloride structures, leading to large population of non-bonding Cl 3\textit{p} orbitals. Third, the ease of local structural distortion around vacancy sites during delithiation promotes dimerization of oxidized Cl anions and ultimately \ch{Cl2} formation. This interpretation is consistent with the much lower mechanical stiffness of chlorides, whose Young’s moduli and hardness are typically one to two orders of magnitude smaller than those of layered oxide cathodes \cite{Fu2025_all_in_one}.

We find that both cation and anion redox potentials can be effectively tuned through halogen substitution, although the shifts in these potentials are strongly correlated. Among the halides considered, fluorides show the most favorable redox characteristics, combining high transition-metal cation redox potentials with an even higher oxidation limit for F–F dimerization. In contrast, Br and I substitutions lower the cation redox potentials relative to those in chlorides, but they also lower the anodic decomposition limit to voltages comparable to or even below the cation redox potentials, thereby limiting the practical utilization of cation redox. We note that partial substitution of Cl by O in \ch{Li2FeCl4} has previously been explored experimentally, where a single phase is obtained up to 5$\%$ O substitution (\ch{Li_{2.2}FeCl_{3.8}O_{0.2}}) \cite{Tanibata2025_redox_exp}. This study shows that the more covalent metal–O bonding lowers the cation redox potential below the Cl--Cl dimerization voltage (similar to the effect of Br and I substitutions demonstrated in our work), expanding the accessible low-voltage capacity. However, it also shows that 5$\%$ O substitution cannot suppress \ch{Cl2} gas release, suggesting that additional mitigation strategies are required to address the Cl--Cl dimerization issue.

When eREAL materials are used as catholytes in composite cathodes, additional design requirements arise regarding electrochemical and mechanical compatibilities with the active material. Both our calculations (Figure \ref{fig:Li2CoCl4_Li2NiCl4_voltage_profiles}) and prior experiments \cite{Liu2024_Li2FeCl4_exp, Fu2025_all_in_one, Wang2023_LTC, Song2024_LVC} show that eREAL materials typically exhibit very flat voltage profiles. Such flat profiles are not ideal for sustaining the mixed-valence states in eREAL catholytes important for robust mixed ionic–electronic conduction. The sloping feature of a voltage profile in oxides is typically associated with solid-solution behavior which is induced by repulsive interactions between the Li ions or by structural disorder. For instance, in cation-disordered rocksalt (DRX) cathodes, cation disorder broadens the distribution of Li site energies, resulting in a steeper slope compared to that of their cation-ordered counterparts \cite{Abdellahi2016_voltage_disordered}. Similarly, it has been demonstrated that introducing cation disorder in oxide spinel \ch{LiMn2O4} can replace the distinct two-phase transition at 3 V with solid-solution behavior \cite{Chen2023_DRX}. However, it is unclear to what extent such structural disorder can enhance the sloping feature in the voltage profiles of eREAL materials. The experimentally measured voltage profile of \ch{Li_{1.6-x}Fe_{1.3}Cl4} is very flat, even though Li and a small amount of Fe are disordered on 4f octahedral sites in the orthorhombic (Cmmm) phase \cite{Fu2025_all_in_one}. Our calculations also indicate that the redox potential is mostly governed by the local [MCl$_6$] octahedral ligand field and local Li site energy, but is less sensitive to the metal-to-Cl ratio, cation ordering, and anion packing. It is likely that Cl anions provide stronger electrostatic screening than O anions due to the larger ionic radius and higher polarizability. This enhanced polarizability effectively screens Li--Li and Li--metal electrostatic interactions, thereby minimizing \ch{Li+}-\ch{Li+} repulsion and site-energy dispersion, suppressing the voltage slope. 

In terms of mechanical compatibility, our calculations suggest that the metal-to-Cl ratio serves as a useful knob for controlling the volume change during cycling: lowering the metal-to-Cl ratio reduces the volume contraction/expansion upon Li extraction/insertion, which is expected to improve mechanical stability during cycling. This benefit, however, comes at the cost of reduced energy density due to a larger fraction of redox-inactive LiCl component within the structure. Ultimately, identifying an optimal eREAL-based cathode materials requires a design strategy that balances electrochemical redox performance, mixed ionic–electronic conduction, and mechanical stability.

\section{Conclusion}

This study highlights the advantages and limitations of eREAL materials for cathode applications. Our first-principles calculations reveal that for most 3d transition metals, a ccp Cl anion framework is favored over an hcp framework due to the small radii of first-row 3\textit{d} transition-metal cations. The strong ionicity of metal--Cl bonds raises the cation redox potentials above those of layered-oxide and olivine cathodes, but also facilitates anion redox due to the limited protection \ch{Cl-} achieves from orbital hybridization. Furthermore, the single negative charge on \ch{Cl-} anions accommodates fewer transition metals than in oxides, leading to more unhybridized Cl 3\textit{p} orbitals in the chloride structure, which further lowers the potential for Cl oxidation. We find that neighboring oxidized Cl anions readily dimerize and evolve \ch{Cl2} at high voltages, constraining the reversible cation-redox capacity. Our results indicate that anion substitution can strongly affect both cation and anion redox potentials, with F substitution emerging as a viable route to extend the cation redox capacity without triggering F oxidation. In agreement with the limited experimental data available, we find that the voltage profile of eREAL materials tends to be flat, which complicates their application as catholytes, because the mixed-valence state important for mixed ionic-electronic conductivity may be difficult to maintain when paired with active materials that operates at distinct voltages or over wider voltage ranges. In summary, we conclude that the flat voltage profiles and the high probability of Cl oxidation and dimerization at elevated potential may be the most challenging issues to introduce these chlorides as redox-active catholytes.

\section*{Author Contributions}
Z.L performed the computations. The manuscript was written by Z.L. and revised by G.C. The work was supervised by G.C.

\section*{Acknowledgment}

This work was supported by the Assistant Secretary for the Office of Critical Minerals and Energy Innovation, Transportation Technologies Office, of the U.S. Department of Energy under Contract No. DE-AC02-05CH11231, under the Advanced Battery Materials Research (BMR) Program. The computational analysis was performed using computational resources sponsored by the Department of Energy’s Office of Energy Efficiency and Renewable Energy located at the National Renewable Energy Laboratory (NREL). Computational resources were also provided by the Advanced Cyberinfrastructure Coordination Ecosystem: Services \& Support (ACCESS) program, which is supported by National Science Foundation grants \#2138259, \#2138286, \#2138307, \#2137603, and \#2138296.

\setstretch{1}
\bibliography{references}

\end{document}